\documentclass{aa}  

\usepackage{graphicx}
\usepackage{txfonts}
\usepackage{natbib}
\usepackage{hyperref}
\usepackage{breakurl}
\newcommand{\mygi}{MyGIsFOS}
\newcommand{\Teff}{\ensuremath{T_\mathrm{eff}}}

\newcommand{\logg}{\ensuremath{\log g}}
\usepackage{color}

\begin{document}

\title{Systematic investigation of chemical abundances derived using IR spectra obtained with GIANO
          \thanks{GIANO programme A31TAC}
}

\author{
E.~Caffau    \inst{1} \and
P.~Bonifacio \inst{1} \and
E.~Oliva     \inst{2}\and
S.~Korotin   \inst{3}\and
L.~Capitanio \inst{1} \and
S.~Andrievsky \inst{4,1}\and
R.~Collet \inst{5}\and
L.~Sbordone \inst{6}\and
S.~Duffau \inst{7}\and
N.~Sanna \inst{2}\and
A.~Tozzi \inst{2}\and
L.~Origlia \inst{8}\and
N.~Ryde \inst{9}\and
H.-G.~Ludwig \inst{10,1}
          }

\institute{GEPI, Observatoire de Paris, Universit\'{e} PSL, CNRS,  5 Place Jules Janssen, 92190 Meudon, France
%              \email{Elisabetta.Caffau@observatoiredeparis.psl.eu}
\and
INAF, Osservatorio Astrofisico di Arcetri, Largo E. Fermi 5, 50125, Firenze, Italy
\and
Crimean Astrophysical Observatory, Nauchny 298409, Republic of Crimea
\and
Astronomical Observatory, Odessa National University, Shevchenko Park, 65014, Odessa, Ukraine
\and
Stellar Astrophysics Centre, Department of Physics and Astronomy, Aarhus University, DK-8000, Aarhus C, Denmark
\and
European Southern Observatory, Casilla 19001, Santiago, Chile
\and
Universidad Andr\'es Bello, Departamento de Ciencias F\'isicas, Fernandez Concha 700. Las Condes, Santiago, Chile
\and
INAF, Osservatorio Astronomico di Bologna, Via Gobetti 93/340129, Bologna, Italy
\and
Lund Observatory, Department of Astronomy and Theoretical Physics, Lund University, Box 43, 221 00 Lund, Sweden
\and
Zentrum f\"ur Astronomie der Universit\"at Heidelberg, Landessternwarte, K\"onigstuhl 12, 69117 Heidelberg, Germany
             }

   \date{Received September 15, 1996; accepted March 16, 1997}

% \abstract{}{}{}{}{} 
% 5 {} token are mandatory
 
  \abstract
  % context heading (optional)
  % {} leave it empty if necessary  
   {Detailed chemical abundances of Galactic stars are needed in order to improve our knowledge of the formation and evolution
of our galaxy, the Milky Way.}
  % aims heading (mandatory)
   {We took  advantage of the GIANO archive spectra to select a sample of Galactic disc stars in order to derive their
chemical inventory and to compare the abundances we derived from these infrared spectra to the
chemical pattern derived from optical spectra.}
  % methods heading (mandatory)
   {We analysed high-quality spectra of 40 stars observed with GIANO.
We derived the stellar parameters from the photometry and the Gaia\,data-release 2 (DR2) parallax;
the chemical abundances were derived with the code MyGIsFOS. For a subsample of stars we compared the
chemical pattern derived from the GIANO spectra with the abundances derived from optical spectra.
We derived P abundances for all 40 stars, increasing the number of Galactic stars for which phosphorus abundance is known.}
  % results heading (mandatory)
   {We could derive abundances of 14 elements, 8 of which are also derived from optical spectra.
The comparison of the abundances derived from infrared and optical spectra is very good.
The chemical pattern of these stars is the one expected for Galactic disc stars and is in agreement with the results from the literature.}
  % conclusions heading (optional), leave it empty if necessary 
   {GIANO is providing the astronomical community with an extremely useful instrument, able to produce
spectra with high resolution and a wide wavelength range in the infrared.}


   \maketitle
%
%-------------------------------------------------------------------

\section{Introduction}

The knowledge of the chemical composition of the stars is a necessary ingredient to understand
and to model the formation and the evolution of the Milky Way and of the Local-Group Galaxies.
In order to derive the chemical pattern of a star, astronomers compare the strengths of the
atomic and molecular lines in an observed stellar spectrum to theoretical predictions.
Typically, spectra with medium/high resolving power (${\rm R}>15000$), a good signal-to-noise ratio
(${\rm S/N}>50$), and a wide-enough wavelength range allow to derive precise abundances for several elements.
Until a few decades ago, securing stellar spectra was the bottleneck in the chemical investigation of stars.
Any single star had to be observed for a reasonable amount of time, depending on its luminosity,
the size of the telescope, and the efficiency of the spectrograph, in order to obtain a 
spectrum with the quality necessary to derive the desired precision in the chemical abundances.
For instance, with a four-meter-class telescope, a spectrum with a typical S/N of about 50 at 520\,nm
is obtained in one hour for a $V = 12$\,mag star \citep[e.g. HARPS][]{Mayor}
while with an eight-meter-class telescope, a spectrum of the same S/N is obtained in one hour 
for a star 300 times fainter in the V band \citep[e.g. FLAMES][]{Flames}.
The advent of the multi-object spectrographs \citep[e.g. FLAMES][]{Flames} allowed one to observe of the
order of one hundred stars at the same time,  considerably increasing the number of stars in the Galaxy for which the chemical composition is known.
The Gaia-ESO Survey \citep{gesgg} is an example of the great efficiency that can be reached by using multi-objects
spectrographs.
In the near future multi-object spectrographs with a capacity to observe about ten times more stars at the same time
are expected (e.g. WEAVE: \citealt{weave} and 4MOST: \citealt{4most}).

Before the commissioning of the near-infrared (NIR) high-resolution spectrograph GIANO \citep{giano14} at 
Telescopio Nazionale Galileo (TNG),
the detailed chemical investigations of stars were based mainly on optical spectra.
Other IR spectrographs appeared before GIANO, such as for example CRIRES \citep[][]{crires04}. 
However, their wavelength coverage was small, useful only to add one element on the chemical 
pattern of a star, and did not allow us to perform a systematic investigation of the IR range.
GIANO is a cross-dispersed echelle spectrograph, and covers a large wavelength range (950-2450\,nm)
allowing the abundance of several elements to be derived.
The multi-object spectrograph  APOGEE \citep{apogee08}, with a wavelength range of about 200\,nm (1510-1700\,nm),
has been shown to be able to provide detailed chemical composition of Galactic stars.
In the  future we expect the Multi Object Optical and Near-infrared Spectrograph \citep[MOONS,][]{moons14}, that will be able to observe
several hundred objects in three arms with two resolution options.
MOONS is a high-multiplicity, multi-object, fibre-fed spectrograph for the ESO VLT.
It will be capable of both low (R$\sim 5\,000$) and high (R$\sim 20\,000$) resolution and will deploy 1000 fibres on the $20'$ field of view of the VLT.
Its high-resolution H arm will observe the wavelength range 1521-1641\,nm, covered by the orders 47-50 of GIANO.

The present work has two main drivers.
On the one hand, we are deeply involved in the preparation of MOONS and wish 
to be prepared for the millions of spectra it will provide. Therefore a detailed study of the IR range is in order.
On the other hand, we are interested in the Galactic evolution of phosphorus, which is still a very little studied  element, 
and of sulphur, both elements being available in the wavelength range of GIANO.
Our main goals in this work are: (i) to investigate the IR range and see which elements can be derived and
(ii) to derive the abundance of phosphorus.
We also wish to provide detailed abundances for several elements for those stars in this sample for which only the abundances
of a few elements are known.
We took the advantage of the GIANO spectra present in the Italian
centre for Astronomical Archives \citep{Molinaro} and we selected a number of dwarf stars (the IR \ion{P}{i} lines are contaminated
in the spectrum of a giant star) of slightly sub-solar metallicity to super-solar metallicity.
For a subsample of stars, SOPHIE \citep{Bouchy} spectra of good quality were available, allowing us to compare for some elements 
the abundances derived from the IR and the optical spectra.

%--------------------------------------------------------------------
\section{Observations}

Stars were observed with GIANO, the high-resolution (R$\simeq$50\,000) 
IR (950--2450 nm) spectrometer of the TNG \citep{giano14}.
The instrument was designed for direct feeding of light at a dedicated
focus of the TNG. 
In 2012 the spectrometer was provisionally positioned on the rotating 
building and fed via a pair of IR (ZBLAN) fibres connected to 
another focal station \citep{tozzi14}. In 2016 it was
eventually moved to the originally foreseen configuration
where it can also be used for simultaneous observations
with HARPS-N \citep{tozzi16}.
The spectra presented here were collected in March 2015, when the 
spectrometer was still in its provisional configuration and fed via fibres.
The raw data were downloaded from the public archive of the TNG.
The stars were observed by nodding on fibre, that is, target and
sky were taken in pairs and alternatively acquired on the two fibers (A,B) 
for an optimal subtraction of the detector noise and background. 
Integration time was 5 minutes per A,B position. Dark-subtracted frames of 
flat (halogen lamp) and wavelength (U-Ne lamp) calibration sources taken
 at the beginning and/or at the
end of each night were used to reduce the data. 
For each nodding pair we created a two-dimensional (2D) frame by subtracting the raw frames (A-B)
and dividing the difference by the normalised flat.
Spectral extraction and wavelength calibration were performed using a
physical model of the spectrometer 
that accurately matches instrumental
effects such as variable slit tilt and orders curvature over the echellogram
\citep{giano2Dreduction}.
The telluric absorption features were corrected using the spectra of two
telluric standards (O-type stars) taken at different airmasses during the 
same nights. The normalised spectra of the telluric standard taken at low
and high airmass values were combined with different weights to match the
depth of the telluric lines in the stellar spectra.

\begin{figure*}
   \centering
   \includegraphics[angle=270,width=18truecm]{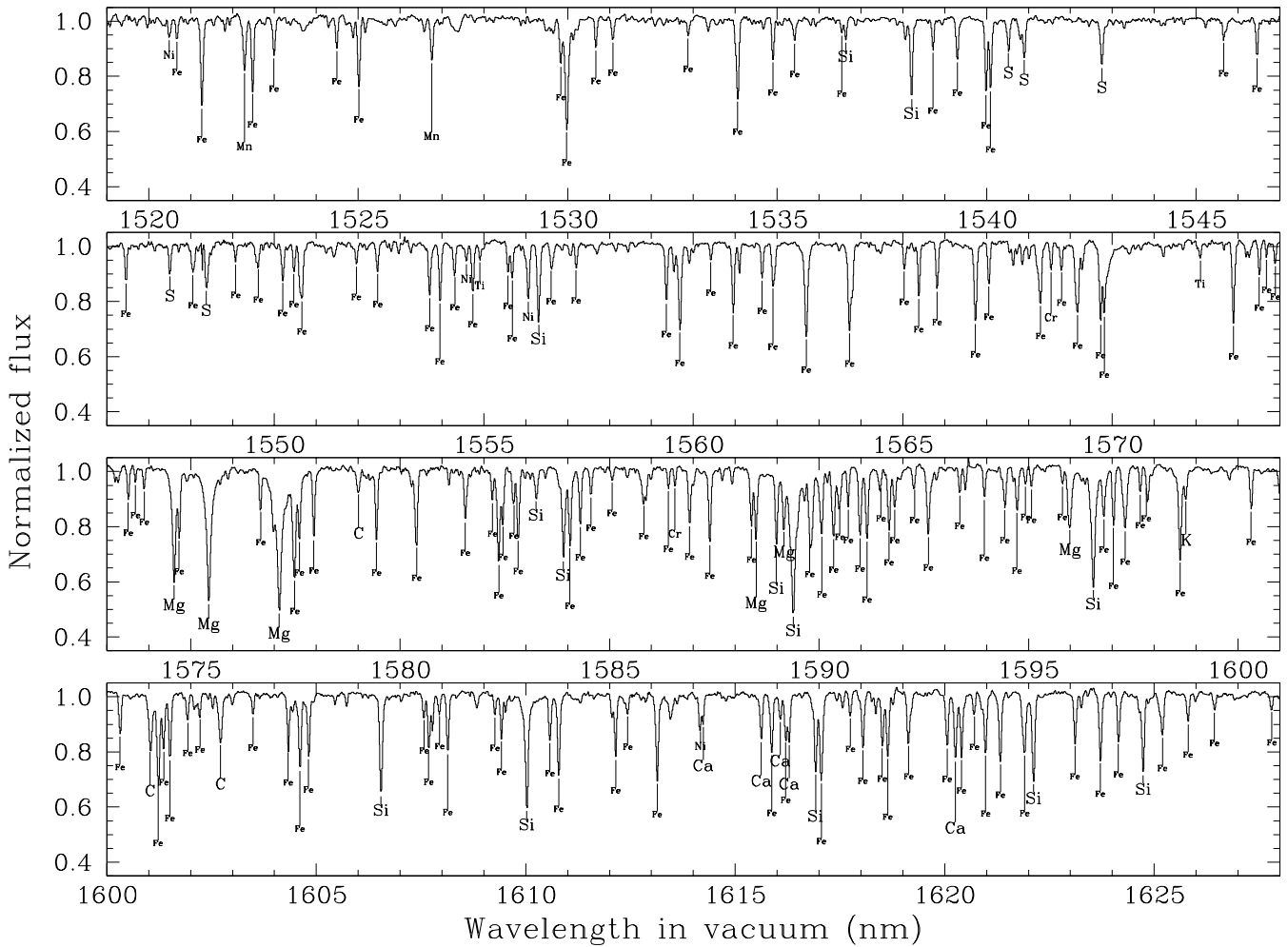}
   \caption{The H-band in the case of HD\,24040 (from top to bottom orders from 50 to 47).}
              \label{o73}%
    \end{figure*}

In Fig.\,\ref{o73} the H-band is shown in the case of the star HD\,24040;
this range corresponds exactly to the wavelength range that will be observed at high resolution by MOONS.

%-------------------------------------- 
%-------------------------------------- 

The SOPHIE spectra have been retrieved from the SOPHIE archive\footnote{\url{http://atlas.obs-hp.fr/sophie/}}.
Several spectra for each star were present in the archive, 
but only the observations with a corresponding sky spectrum were retained.
We subtracted the sky, shifted the spectra at laboratory
wavelength and, in the case of several observations, added them together.
For one star, HD\,76909, we had a single observation.

%%%%%%%%%%%%%%%%%%%%%%%%%%%%%%%%%%%%%%%%%%%%%%%%%%%%%%%%%%%%%%%%%%%%%%%%
\section{Stellar parameters}

Within the stars observed by GIANO available in the archive
we selected those which had Gaia DR2  \citep{DR2,gaiababusiaux}  parallax as  well as
2MASS \citep{cutri} and Tycho \citep{1997ESASP1200.....E}  photometry.
To derive effective temperatures we used the colour-temperature
calibrations of \citet{IRFM09} that are based on temperatures
derived through the IR flux method for a sample of calibrators.
To do so we used the colour transformations given by equations 
1.3.20 and 1.3.26 \relax in the Hipparcos and Tycho catalogues 
\citep{1997ESASP1200.....E}\footnote{\begin{eqnarray*}
V_J = V_T - 0.090 *(B-V)_T \\
  z = B_T-V_T -0.90 \\
(B-V)_J =  (B_T-V_T) -0.115 -0.229 z +0.043 z^3
\end{eqnarray*}}
%\citep{1997ESASP1200.....E}\footnote{
%$V_J = V_T - 0.090 *(B-V)_T$ \\
%$z = B_T-V_T -0.90$ \\
%{1truecm}  $(B-V)_J =  (B_T-V_T) -0.115 -0.229 z +0.043 z^3$}
to obtain $V$ an $(B-V)$ in the Johnson system.
All the stars are fairly near and little reddened, nevertheless
we derived the reddening for each star from the reddening
maps of \citet{Capitanio} and corrected the colours for reddening.
We then applied the equations given
in table 5 of \citet{IRFM09} to 
derive effective temperatures (\Teff) from $(B-V)$, $(V-J)$, $(V-H)$, $(V-J)$ and $(V-K)$ that were averaged 
among the different calibrations. The colours involving the 2MASS photometry were retained
only if the corresponding quality flag was A. This implies that not all stars have the same
number of colour determinations.
In Table\,\ref{fevalues}, \Teff\ is provided with an uncertainty which is the scatter among the temperatures derived by the five calibrations.
For one star, HD\,190360, \Teff\ is derived from one single calibration (that using the $B-V$ colour), and  we assumed 250\,K
as uncertainty in the temperature; for two stars, HD\,28005 and HD\,34575, four calibrations were used
(those based on colours $B-V, V-J,V-K$ and $J-K$),
 and for all the other stars the five calibrations were used.
The effective temperature of this sample of stars spans from 5006\,K to 6292\,K, with
$\langle{{\rm T}_{\rm eff}}\rangle = \left(5727\pm 245\right)$\,K.

For each calibration used to derive the effective temperature, we derived the 
corresponding surface gravity
from the Stefan-Boltzmann relation using the distances from the parallaxes provided by Gaia DR2 and the stellar absolute luminosity,
derived from the Gaia $G$ band and a bolometric correction based on three-dimensional (3D) model atmospheres
\citep{Bonifacio18}. We assumed the mass of each star according to its temperature, by looking at the isochrones.
The uncertainty in the gravity is the scatter among the values we derived, 
and this value is on average below 0.05 and for all stars below 0.1.
The gravities we derived spanned from 3.53 to 4.56 [c.g.s], with $\langle{\log {\rm g}}\rangle = 4.16\pm 0.25$
(see Table\,\ref{fevalues}).

The stars we are analysing have a metallicity from slightly sub-solar to super-solar, 
meaning that the lines are strong and several lines are blended.
Also, to derive the micro-turbulence in the IR range is not trivial since the lines we can detect are stronger
than in the optical range.
We then fixed the value for the microturbulence at 1.0\,km/s, which is a good value for this sample of dwarf stars.
Also for the Sun a microturbulence at 1.0\,km/s is the preferred value \citep{steffen13}.

The chemical analysis has been done with the pipeline \mygi\ \citep{mygi14}, in the complete wavelength range
provided by GIANO, except the ranges that are too heavily contaminated by telluric absorption.
This choice was also motivated by two facts:
(i) the GIANO archive is providing more and more stars that can be analysed; (ii) the future MOONS-GTO as well as the future MOONS observations
will imply several thousand  spectra each night. The numbers of stars to analyse is too large to reasonably allow a by-hand analysis. 
To analyse the present sample of stars, \mygi\ operated at fixed \Teff\ and \logg\ and derived, from 
selected atomic lines, the chemical abundances by comparing the observed spectrum to a grid of synthetic spectra.
The theoretical synthesis was computed by SYNTHE \citep[see][]{kurucz05,sbordone04} 
from a grid of ATLAS\,12 models \citep{kurucz05} with a step of 0.25\,dex in metallicity.
The atomic parameters of the lines are from the most recent line-list provided by R. Kurucz on his 
site\footnote{\url{http://kurucz.harvard.edu/linelists.html}}; in the H band we used the atomic data provided by \citet{apogeell15}.
The stars we analysed all had metallicities from slightly sub-solar to super-solar, and from \ion{Fe}{i} lines we derived for the sample of stars:
$\langle{[{\rm Fe/H}]}\rangle = +0.10\pm 0.11$;
for all the stars $-0.14\leq[{\rm Fe/H}]\leq +0.33$.
Similar results were derived using \ion{Fe}{ii} lines, $-0.07\leq[{\rm Fe/H}]\leq +0.40$,
with an average difference of [Fe/H] from \ion{Fe}{i} and \ion{Fe}{ii} lines smaller than 0.01\,dex,
 the largest one being just smaller than 0.10\,dex. 
The difference of [Fe/H] from \ion{Fe}{i} and \ion{Fe}{ii} lines is in any case smaller than the uncertainties
(see Table\,\ref{fevalues}).

%%%%%%%%%%%%%%%%%%%%%%%%%%%%%%%%%%%%%%%%%%%%%%%%%%%%%%%%%%%%%%%%%%%%%%%%
\section{Comparison optical - IR}

For nine stars (see Table\,\ref{nines}), we had optical SOPHIE spectra for which one fibre was on the sky.
We could then subtract the sky and compare the analysis of GIANO and SOPHIE for
the elements in common.

\begin{figure}
   \centering
   \includegraphics[width=\hsize]{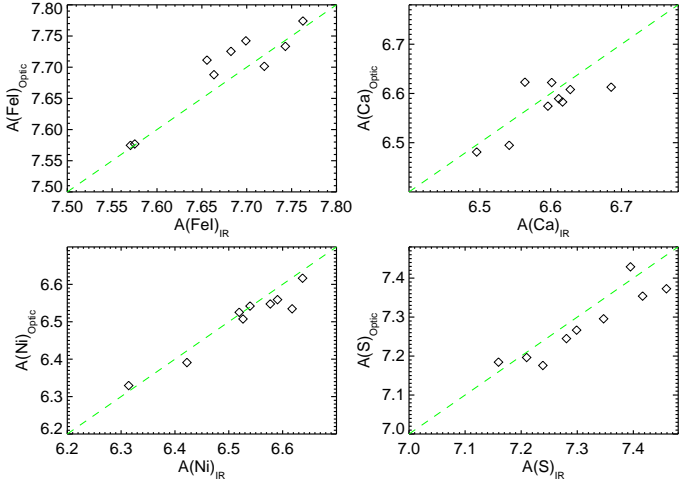}
      \caption{Comparison of the abundances derived from IR and optical spectra.}
         \label{compabbo}
   \end{figure}

Eight elements are in common in the two analyses and the agreement is generally very good.
\begin{itemize}
%Fe
\item{ }The agreement in the Fe abundance derived from optical and IR spectra is good:
$\langle{\rm A(\ion{Fe}{i})}_{\rm IR}-{\rm A(\ion{Fe}{i})}_{\rm opt}\rangle = -0.02\pm 0.02$ (see Fig.\,\ref{compabbo})
and $\langle{\rm A(\ion{Fe}{ii})}_{\rm IR}-{\rm A(\ion{Fe}{ii})}_{\rm opt}\rangle = 0.03\pm 0.09$.
For one star in particular (HD\,76909), the difference on A(\ion{Fe}{ii}) from optical and IR spectra
is large (0.18\,dex) but still within uncertainties; in fact it is the optical spectrum that shows a larger disagreement in the A(Fe)
from \ion{Fe}{i} and \ion{Fe}{ii} lines and the line-to-line scatter from the \ion{Fe}{ii} lines is the largest
among the optical spectra.
%Na
\item{ }There is a systematic difference in A(Na) from IR and optical spectra:
$\langle{\rm A(Na)}_{\rm IR}-{\rm A(Na)}_{\rm opt}\rangle = -0.23\pm 0.14$.
From the IR and optical spectra we have $\langle[{\rm Na/Fe}]\rangle = -0.11\pm 0.09$
and $\langle[{\rm Na/Fe}]\rangle = +0.11\pm 0.08$, respectively.
This difference could  be due to NLTE effects; in fact the IR lines form close to LTE, but
this is not true for the optical lines, which are sensitive to NLTE. 
%Al
\item{ }The agreement in A(Al) is really good; in the case of IR spectra, in seven cases out of nine, A(Al) is
derived from one single line and in the other two stars from two lines: 
$\langle{\rm A(Al)}_{\rm IR}-{\rm A(Al)}_{\rm opt}\rangle = +0.07\pm 0.06$.
%Si
\item{ }For A(Si) from \ion{Si}{i} lines, the agreement of IR and optical spectra is good, well within the uncertainties:
$\langle{\rm A(Si)}_{\rm IR}-{\rm A(Si)}_{\rm opt}\rangle = -0.08\pm 0.04$.
%S
\item{ }Also the agreement for A(S) is extremely good:
$\langle{\rm A(S)}_{\rm IR}-{\rm A(S)}_{\rm opt}\rangle = +0.03\pm 0.04$
(see Fig.\,\ref{compabbo}).
%Ca
\item{ }For A(Ca) from \ion{Ca}{i} lines the agreement is also perfect:
$\langle{\rm A(Ca)}_{\rm IR}-{\rm A(Ca)}_{\rm opt}\rangle = +0.02\pm 0.04$
(see Fig.\,\ref{compabbo}).
%Ti
\item{ }The Ti abundance, derived from \ion{Ti}{i} and \ion{Ti}{ii} lines is very close from IR and optical spectra:
$\langle{\rm A(\ion{Ti}{i})}_{\rm IR}-{\rm A(\ion{Ti}{i})}_{\rm opt}\rangle = -0.05\pm 0.05$ and 
$\langle{\rm A(\ion{Ti}{ii})}_{\rm IR}-{\rm A(\ion{Ti}{ii})}_{\rm opt}\rangle = 0.08\pm 0.05$.
%Ni
\item{ }For Ni we also find a good agreement between IR and optical:
$\langle{\rm A(Ni)}_{\rm IR}-{\rm A(Ni)}_{\rm opt}\rangle = +0.02\pm 0.03$
(see Fig.\,\ref{compabbo}).
\end{itemize}

A summary on the comparison of the abundances derived from IR and
optical spectra is presented in Table\,\ref{IR_vis_el}.

%-------------------------------------------------------------
%
%-------------------------------------------------------------

%%%%%%%%%%%%%%%%%%%%%%%%%%%% Table %%%%%%%%%%%%%%%%%%%%%%%%%%%%%%%%%%%%%
\begin{table}
\caption{Sample of stars with optical and IR spectra.}
\label{nines}
\renewcommand{\tabcolsep}{3pt}
\tabskip=1pt
\begin{center}
\begin{tabular}{lrrrrrrrrrrrrr}
\hline\hline{\smallskip}
{Star}   & {\Teff} & \logg  & $\xi$ & \multicolumn{2}{c}{[Fe/H]} \\
         &   K     &  [cgs] & km/s  & GIANO & Optical \\
\hline \noalign{\smallskip}
HD\,34575  &  5582 & 4.22 & 1.0 &  0.18 &  0.22 \\ 
HD\,67346  &  5953 & 3.78 & 1.0 &  0.14 &  0.19 \\
HD\,69056  &  5637 & 4.28 & 1.0 &  0.05 &  0.05 \\ 
HD\,69960  &  5655 & 3.99 & 1.0 &  0.22 &  0.21 \\ 
HD\,73933  &  6143 & 4.24 & 1.0 &  0.06 &  0.06 \\
HD\,76909  &  5655 & 4.17 & 1.0 &  0.24 &  0.25 \\ 
HD\,90681  &  5950 & 4.26 & 1.0 &  0.16 &  0.21 \\
HD\,97645  &  6127 & 4.10 & 1.0 &  0.14 &  0.17 \\ 
HD\,108942 &  5882 & 4.27 & 1.0 &  0.20 &  0.18 \\ 
\noalign{\smallskip}
\hline
\hline
\end{tabular}
\end{center}
\end{table}
%%%%%%%%%%%%%%%%%%%%%%%%%%%%%%%%%%%%%%%%%%%%%%%%%%%%%%%%%%%%%%%%%%%%%%%%

%%%%%%%%%%%%%%%%%%%%%%%%%%%% Table %%%%%%%%%%%%%%%%%%%%%%%%%%%%%%%%%%%%%
\begin{table}
\caption{ Comparison of abundances from IR and optical spectra.}
\label{IR_vis_el}
\renewcommand{\tabcolsep}{3pt}
\tabskip=1pt
\begin{center}
\begin{tabular}{lrr}
\hline\hline{\smallskip}
{Ion}   & IR--optical & r.m.s. \\
\hline \noalign{\smallskip}
\ion{Na}{i}  & $-0.23$ & 0.14 \\
\ion{Al}{i}  & $+0.07$ & 0.06 \\
\ion{Si}{i}  & $-0.08$ & 0.04 \\
\ion{S}{i}   & $+0.03$ & 0.04 \\
\ion{Ca}{i}  & $+0.02$ & 0.04 \\
\ion{Ti}{i}  & $-0.05$ & 0.05 \\
\ion{Ti}{ii} & $+0.08$ & 0.05 \\
\ion{Fe}{i}  & $-0.02$ & 0.02 \\
\ion{Fe}{ii} & $+0.03$ & 0.09 \\
\ion{Ni}{i}  & $+0.02$ & 0.03 \\
\noalign{\smallskip}
\hline
\hline
\end{tabular}
\end{center}
\end{table}
%%%%%%%%%%%%%%%%%%%%%%%%%%%%%%%%%%%%%%%%%%%%%%%%%%%%%%%%%%%%%%%%%%%%%%%%

Previously published investigations exist for all nine stars from Table 1.
Three of the papers that contain parameters for some of these stars
are based on spectra \citep{valenti05,petigura11,brewer16}
and two are based on photometry \citep{masana06,casagrande11}.
All the  spectra used in %\LEt{please only use semicolons to separate references inside parentheses; regular punctuation should be used in all examples like this where references are used in the body text; please modify throughout where appropriate.} 
\citet{valenti05,petigura11,brewer16} are optical;
all have been collected in the course of planet search campaigns and have
S/N ratios exceeding 100.
\citet{valenti05} used three different spectrographs (HIRES at Keck, 
Hamilton at Lick, and UCLES at the AAT), mostly
with a resolution of $R\sim 70\,000$, comparable to our SOPHIE
spectra. Also \citet{petigura11} used HIRES spectra, but at a somewhat lower
resolution, $R\sim 50\,000$. Finally \citet{brewer16}
used HIRES at Keck, and used several resolutions 
in the range $57\,000 \le R \la 78\,000$. 
In the following we discuss abundances in terms of A(X) = log(X/H)+12.

%%%%%%%%%%%%%%%%%%%%%%%
\subsection{HD\,34575}
For this star we derived ${\Teff} = 5582$\,K, in very good agreement with 5529\,K derived by \citet{masana06}
and 5546\,K derived by \citet{casagrande11}.
The analysis by \citet{brewer16} is in very good agreement with our investigation;
they derive very close stellar parameters (${\Teff} = 5551$\,K, ${\rm\logg} = 4.25$) and also the abundances 
(C, Na, Al, Si, Ca, Ti, Fe) are
in agreement with our findings, well within the uncertainties.
 \citet{valenti05} derived a close effective temperature, 5651\,K, and a gravity 0.23\,dex larger than our adopted  value. 
The abundances they derive (Na, Si, Ti, Fe, Ni) are mostly slightly larger, but well compatible within the uncertainties.
With the stellar parameters from \citet{valenti05}, \citet{petigura11} derived a C abundance for this star that is
compatible with our value, but larger, as in the abundances from \citet{valenti05}.

%%%%%%%%%%%%%%%%%%%%%%%
\subsection{HD\,67346}
The temperature we derive for this star of 5953\,K is in close agreement with ${\Teff} = 5941$\,K from \citet{masana06},
but 200\,K cooler than \citet{casagrande11} (${\Teff} = 6157$\,K).
With a \Teff\ that is 140\,K hotter and a gravity of 0.4\,dex higher, \citet{petigura11} derived higher C and Ni abundances.

%%%%%%%%%%%%%%%%%%%%%%%
\subsection{HD\,69056}
The effective temperature we derive ($\Teff = 5637$\,K) for this star is in very good agreement with the values from
\citet{masana06} and \citet{casagrande11} (\Teff\ of 5598 and 5635\,K, respectively).
\citet{petigura11} used $\Teff = 5490$\,K and \logg\ of 4.27 and derived C and Ni abundances in agreement with our results.
%The star shows an overabundance in Sr.

%%%%%%%%%%%%%%%%%%%%%%%
\subsection{HD\,69960}
Our $\Teff = 5655$\,K is in extremely good agreement with $\Teff = 5625$\,K by \citet{casagrande11} and
also in agreement with the cooler value of \citet{masana06} ($\Teff = 5572$\,K).
The gravity we derived is of 3.99.
\citet{brewer16} with very similar stellar parameters, derived abundances (C, Na, Mg, Al, Si, Ca, Ti Cr, Fe, Ni) 
that are in general good agreement within uncertainties, with our results.
With stellar parameters of $\Teff = 5690$\,K and $\logg = 4.21$, \citet{petigura11} derive ${\rm A(C)} = 8.76$, consistent
with our value within uncertainties. 

We are, by and large, in agreement with \citet{chen08}, in spite of their effective
temperature being about 200\,K cooler.
%We would also be in general agreement with \citet{chen08} if their effective temperature were about 200\,K cooler.
%\LEt{Please check that I have retained your intended meaning.}

%%%%%%%%%%%%%%%%%%%%%%%
\subsection{HD\,73933}
The effective temperature we derive (6243\,K) is about 100\,K hotter than the values from  \citet{casagrande11} and \citet{masana06}.
\citet{petigura11}, with an effective temperature of 6076\,K and $\logg = 4.39$,
derived a C abundance in agreement with our value, within uncertainties

%%%%%%%%%%%%%%%%%%%%%%%
\subsection{HD\,76909}
The $\Teff = 5655$\,K we adopt is in very good agreement with \citet{masana06}.
\citet{valenti05} adopted an effective temperature about 100\,K higher, a gravity 0.2\,dex higher,
and derive a metallicity about 0.1\,dex higher.
Our results are in very good agreement with those of \citet{brewer16} both for the stellar parameters and the detailed abundances
(C, Mg, Al, Si, Ca, Ti, Cr, Fe, Ni). Only our Na abundance is definitely larger.
\citet{petigura11} adopted parameters compatible with ours ($\Teff = 5740$\,K and $\logg = 4.39$)
to derive a C abundance in perfect agreement with our value.

We remind the reader that we used
only one SOPHIE spectrum for the optical analysis, the only one that has the sky.

%%%%%%%%%%%%%%%%%%%%%%%
\subsection{HD\,90681}
For this star we adopted $\Teff = 5950$\,K and $\logg = 4.26$.
With very close stellar parameters ($\Teff = 5995$\,K and $\logg = 4.49$), \citet{petigura11} derived the same C abundance.

%%%%%%%%%%%%%%%%%%%%%%%
\subsection{HD\,97645}
We adopted $\Teff = 6127$\,K and $\logg = 4.10$.
With consistent stellar parameters ($\Teff = 6171$\,K and $\logg = 4.32$), \citet{petigura11} derived a 
C abundance (A(C)=8.51) very close to our value of A(C)=8.48.

%%%%%%%%%%%%%%%%%%%%%%%
\subsection{HD\,108942}
We adopted stellar parameters ($\Teff = 5882$\,K and $\logg = 4.27$) in good agreement with the literature
\citep{masana06,casagrande11,ramirez12}.
\citet{petigura11}, adopting a slightly smaller $\Teff = 5795$\,K and higher $\logg = 4.32$,
derived abundances (C, Na, Mg, Al, Si, Ti, Cr, Fe, Ni) in agreement, within uncertainties, with our values.

%%%%%%%%%%%%%%%%%%%%%%%%%%%%%%%%%%%%%%%%%%%%%%%%%%%%%%%%%%%%%%%%%%%%%%%%%
\subsection{Granulation effects}

\begin{figure}
   \centering
   \includegraphics[width=\columnwidth]{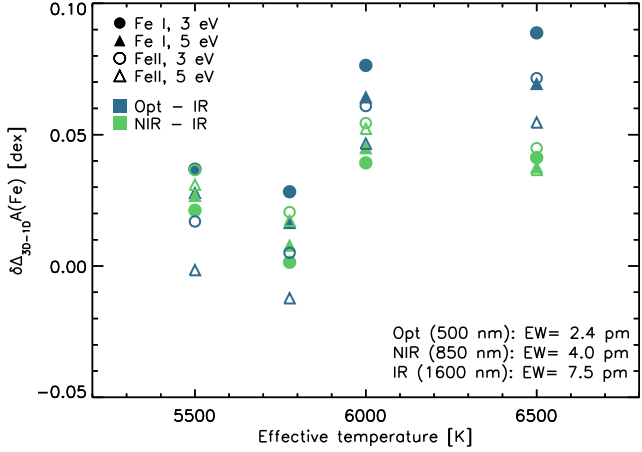}
   \caption{Differential granulation effects, i.e. differential {3D$-$1D} abundance corrections 
$\delta\,\Delta_{\rm 3D-1D}\,A({\rm Fe})$, between optical and IR and between NIR and IR wavelengths 
derived from moderately-high-excitation unsaturated Fe lines for dwarf stars as a function of effective temperature. 
Filled and open symbols refer to \ion{Fe}{i} and \ion{Fe}{ii} lines, respectively. 
Circles and triangles represent lines with a lower-level excitation potential of $3$ and $5$\,eV, respectively. 
Blue (green) symbols indicate differential corrections between optical and IR (NIR and IR) wavelengths. 
Abundance corrections are computed for fictitious lines with the same normalised equivalent width, ${\rm EW}/\lambda,$ 
and corresponding to a value of $w_\lambda = 4$\,{pm} at $\lambda = 850$\,nm.}
   \label{fig:diff_corr}
\end{figure}

We investigated whether the effects of granulation on abundance determinations are similar at optical and IR wavelengths.
In order to assess this, we synthesised a number of fictitious \ion{Fe}{i} and \ion{Fe}{ii} lines 
of varying strength with excitation potentials between $3$ and $5$\,eV at $500$\,nm (optical), $850$\,nm (NIR), and $1\,600$\,nm (IR) 
using both 1D and 3D model atmospheres of solar-metallicity 
(${\rm [Fe/H]} = +0.0$) for stars with physical parameters representative of the present sample: 
\Teff\ in the range $5\,000-6\,500$~K and $\log{g} = 4.0$~(cm\,s$^{-2}$) plus the Sun 
($\Teff = 5771$\,K and $\log{g} = 4.44$).
The 3D hydrodynamic and corresponding stationary 1D hydrostatic models 
were adopted from the {\sc stagger-grid} \citep{Magic:2013}.
We used these models in combination with the {\sc scate} code \citep{Hayek:2011} to compute spectral line 
profiles under the assumption of local thermodynamic equilibrium (LTE).
For the calculations with 1D models, we adopted a micro-turbulent parameter of $\xi = 1.0$~km\,s$^{-1}$.
We derived the {3D$-$1D} abundance corrections for the fictitious lines with varying strength by comparing the 
3D and 1D Fe abundances required to match a given equivalent width (EW) value.
We then quantified the differential granulation effects between different spectral regions by computing 
the  {3D--1D} abundance corrections of lines from the same ion at  two different wavelengths, 
for example,  
optical and IR (but with identical excitation potential and identical normalised equivalent width ${\rm EW}/\lambda$) and analysing the difference.
%\LEt{Please check that I have retained your intended meaning.}.
We find the differential granulation effects for unsaturated \ion{Fe}{ii} lines at optical and IR wavelengths 
to be generally small, ranging from about $0.0$~dex at $\Teff = 5500$\,K to about $0.06$\,dex at $\Teff = 6500$\,K. 
Differential granulation effects for \ion{Fe}{i} lines are typically 0.01 to 0.02\,dex larger, but following a similar 
trend with effective temperature. For comparison, differential {3D$-$1D} abundance corrections between NIR and IR 
lines show a flatter trend with values in the range $0.00-0.05$\,dex. Results are summarised in Fig.\,\ref{fig:diff_corr}. 
Calculations performed with models from the {\sc CIFIST} grid \citep{cifist} show similar behaviour for 
the differential granulation effects \citep[see also][]{zolfito}.

%%%%%%%%%%%%%%%%%%%%%%%%%%%%%%%%%%%%%%%%%%%%%%%%%%%%%%%%%%%%%%%%%%%%%%%%%
\section{Detailed chemical investigation of the full sample}

We were able to derive the abundance of 14 elements (C, Na, Mg, Al, Si, P, S, K, Ca, Ti, Cr, Fe, Ni, Sr) from the GIANO spectra. These are
reported in Tables\,\ref{fevalues}-\ref{abbocrnisr}. 
Some lines of \ion{Mn}{i} are present in the range of GIANO (1297, 1328, 1329, 1521\,nm), but the
abundance we derive is not trustworthy, probably due to departure from LTE or granulation effects.

%%%%%%%%%%%%%%%%%%%%%%%
\subsection{Phosphorus}

Phosphorus has a single stable isotope $^{31}$P, which is probably produced via neutron capture
and the most likely site where it is produced is in the oxygen and neon burning shells during late stages of massive stars
\citep[see e.g.][]{phosphoro2011}.

\begin{figure}
   \centering
   \includegraphics[width=\hsize,clip=true]{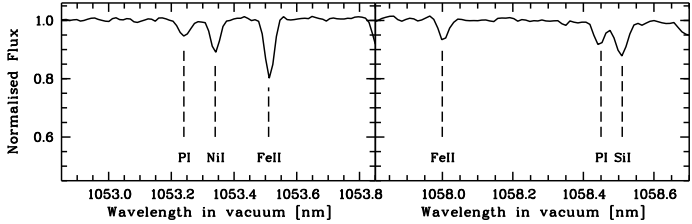}
      \caption{The two lines of \ion{P}{i} belonging to  Mult.1 for the star HD\,24040.}
         \label{hd24040pi}
   \end{figure}

With this analysis, we added 40 points in the [P/Fe] versus [Fe/H] diagram.
The \ion{P}{i} lines we looked at are two of the lines of Mult.\,1 at 1053.24 and 1058.44\,nm 
($\log{\rm gf}$ of 0.20 and 0.48, respectively),
which have already been used by \citet{phosphoro2011,phosphoro2016},
and a line of Mult.\,4 at 1008.70\,nm (with $\log{\rm gf} = 0.130$).
The two lines of Mult.\,1 are shown in Fig.\,\ref{hd24040pi} in the case of HD\,24040.
The P abundance we derive from the 1008.70\,nm line is on average larger (0.06 and 0.15\,dex) 
than the one derived from the other two lines.
This difference is not particularly large,  the difference between the two lines of Mult.\,1 being 0.09\,dex.

\begin{figure}
   \centering
   \includegraphics[width=\hsize,clip=true]{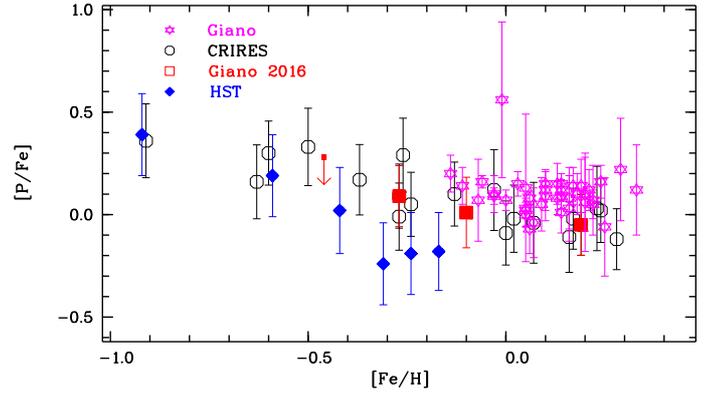}
      \caption{[P/Fe] vs. [Fe/H] is shown for the programme stars (pink stars).
The four stars analysed by \citet{phosphoro2016} are the solid red squares, the results from \citet{phosphoro2011} are the black open circles, and 
the most metal-poor sample is from \citet{roederer14} (solid blue diamonds).}
         \label{pabbo}
   \end{figure}

In Fig.\ref{pabbo}, we compare these results with previous analyses.
In particular, \citep{phosphoro2011,phosphoro2016} claimed a behaviour of P similar
to an $\alpha$-element, such as S. Unfortunately, the metallicity of this sample
of stars is too high to be able to confirm this finding.

%-------------------------------------------------------------
%-------------------------------------------------------------

%-------------------------------------------------------------
%-------------------------------------------------------------

%%%%%%%%%%%%%%%%%%%%
\subsection{Sulphur}

Sulphur is an $\alpha$ element produced by type-II supernovae.
The fact that sulphur is a relatively volatile element, preventing it from being locked into dust grains in the 
interstellar medium, makes it a perfect element to compare $\alpha$ abundances in stars and 
in blue compact galaxies and damped Ly$\alpha$ systems.
Details on S and an exhaustive summary of the results in the literature can be found in \citet{duffau17}.

\begin{figure}
   \centering
   \includegraphics[width=\hsize,clip=true]{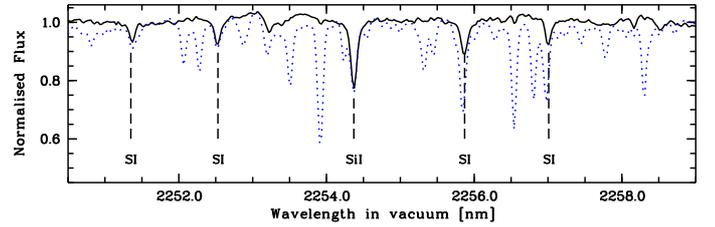}
      \caption{Four of the \ion{S}{i} lines belonging to a multiplet (solid black) in the case of the star HD\,24040.
 The spectrum (dotted blue) before the correction of the telluric lines is overplotted.
}
         \label{s2261hd24040}
   \end{figure}

To derive A(S), we investigated the three \ion{S}{i} lines of Mult.\,3 at 1045\,nm as we did in \citet{phosphoro2016}.
We also looked at the triplet at 1541\,nm, 
one single line in the multiplet at 1547\,nm and 
the multiplet at 2261\,nm (see Fig.\,\ref{s2261hd24040} in the case of HD\,24040). 
This multiplet is in a wavelength range contaminated by telluric absorptions that
have to be removed when it is analysed.

\begin{figure}
   \centering
   \includegraphics[width=\hsize,clip=true]{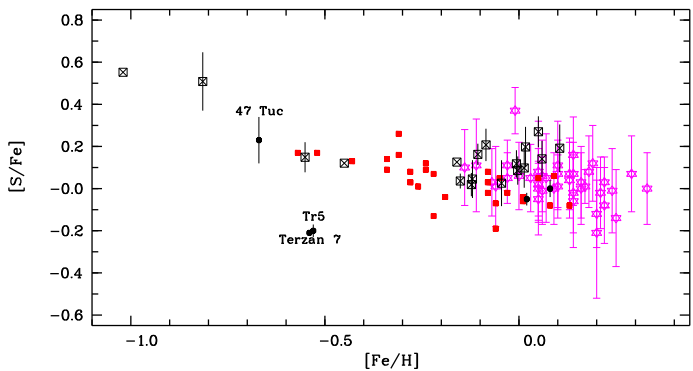}
      \caption{An update of the plot shown by \citet{duffau17} in its metal-rich range, for [S/Fe] vs. [Fe/H].
The program stars are shown as pink stars.
The red filled squares are from \citet{ecuvillon04}.
}
         \label{sabbo}
   \end{figure}

We can compare the S abundance derived here with the metal-rich sample analysed by \citet{ecuvillon04}
and with our previous investigations on sulphur.
In Fig.\ref{sabbo}, we added the sample  analysed here to the metal-rich range of the plot we showed in \citet{duffau17};
the agreement with the previous investigations is very good.
The fact that [S/Fe] analysed here is on average slightly higher than the result by \citet{ecuvillon04}
could be explained by the fact that among the \ion{S}{i} lines that we adopt to derive the S abundance,
the lines of Mult.\,3 at 1045\,nm are affected by NLTE, which would slightly decrease A(S), while \citet{ecuvillon04}
analysed the lines of Mult.\,8 formed close to the LTE condition.

%-------------------------------------------------------------
%-------------------------------------------------------------

%-------------------------------------------------------------
%-------------------------------------------------------------

%%%%%%%%%%%%%%%%%%%%%%%%%%%%%%%
\subsection{The other elements}

 We were able to find several atomic lines of \ion{C}{i }in the GIANO wavelength   range.
This is extremely important because nitrogen can be derived from CN molecules
once the C abundance is known.
[C/Fe] is consistent with Galactic stars of slightly sub-solar to super-solar metallicity
($[{\rm C/Fe}] = -0.04\pm 0.08$) with just one star, HD\,190360, with $[{\rm C/Fe}]>+0.3$.

\begin{figure}
   \centering
   \includegraphics[width=\hsize,clip=true]{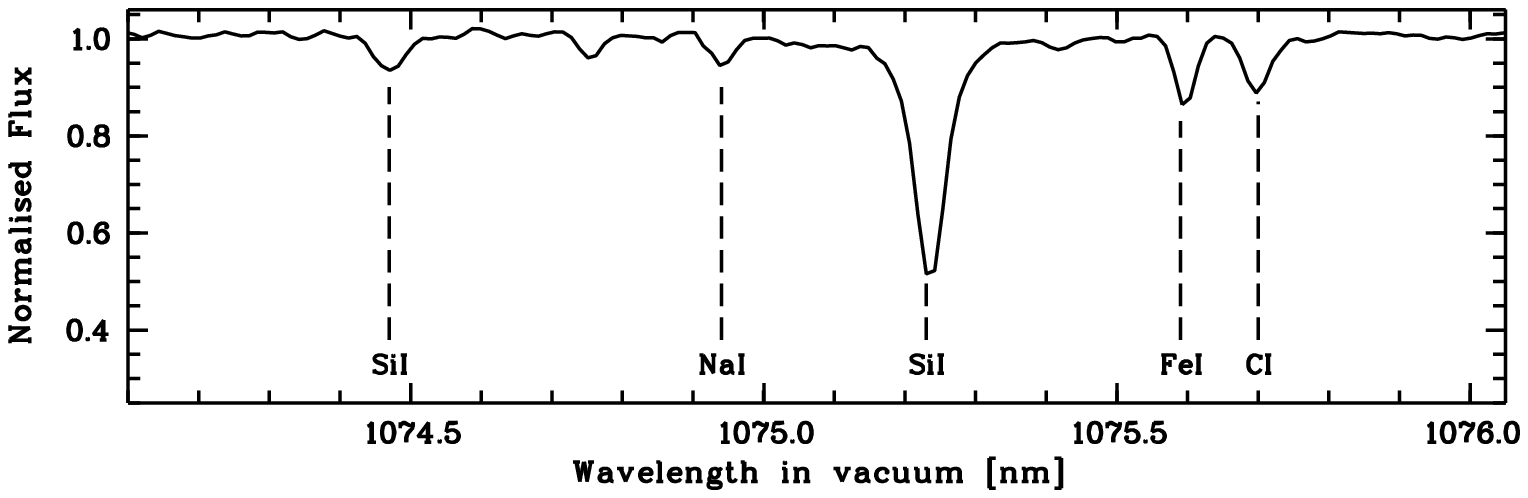}
      \caption{The \ion{Na}{i} line at 1075\,nm (solid black) in the case of the star HD\,24040.
}
         \label{na1075hd24040}
   \end{figure}

We used three lines of \ion{Na}{i} to derive the Na abundances; according to our calculations,
they all  form close to LTE and therefore we do not expect large NLTE corrections
(see in Fig.\,\ref{na1075hd24040} the line at 1075\,nm for HD\,24040).
The NLTE atomic model of sodium was presented by \citet{korotinmishenina99}
and then updated by \citet{dobrovolskas14}. The updated sodium model currently consists of twenty energy levels of \ion{Na}{i} and the
ground level of \ion{Na}{ii}. In total, 46 radiative transitions were taken
into account for the calculation of the population of all levels.
We carried out test calculations for a grid of stellar atmosphere
models in the ranges: $5000 < \Teff <6000$\,K and $3.0<\logg <4.5$ with solar metallicity. 
For the IR sodium lines, NLTE effects lead to a slight increase in EW by 1 to 3\%, while the EW in the optical lines
increases by 10-18\%.
For all but one star we were able to derive the Na abundance with $\langle{\rm [Na/Fe]}\rangle = +0.05\pm 0.17$ in the sample.
The large scatter is due to six stars with [Na/Fe] around +0.1\,dex
and several stars with negative [Na/Fe], down to about $-0.2$\,dex.

The \ion{Mg}{i} lines we selected to derive the Mg abundance are formed close to the LTE condition.
In the case of Mg, we used the model atom from \citet{mishenina04}, then updated by \citet{cerniauskas17}. 
It consisted of 84 levels of \ion{Mg}{i}, 12 levels of \ion{Mg}{ii}, and the ground
state of \ion{Mg}{iii}. In the computation of departure coefficients,
the radiative transitions between the first 59 levels of \ion{Mg}{i} and the ground level of \ion{Mg}{ii} were taken into account.
We carried out test calculations within the same grid of star
parameters as for sodium. It turned out that due to the influence of
the effects of the deviation from the LTE, the EW of the magnesium lines used in this analysis vary by $\pm 1$\%; that is, these lines are formed almost in the LTE.
The average [Mg/Fe] we derived ($\langle{\rm [Mg/Fe]}\rangle = +0.02\pm 0.07$)
is the one expected for disc stars around solar metallicity.

%-------------------------------------------------------------
%-------------------------------------------------------------

\begin{figure}
   \centering
   \includegraphics[width=\hsize,clip=true]{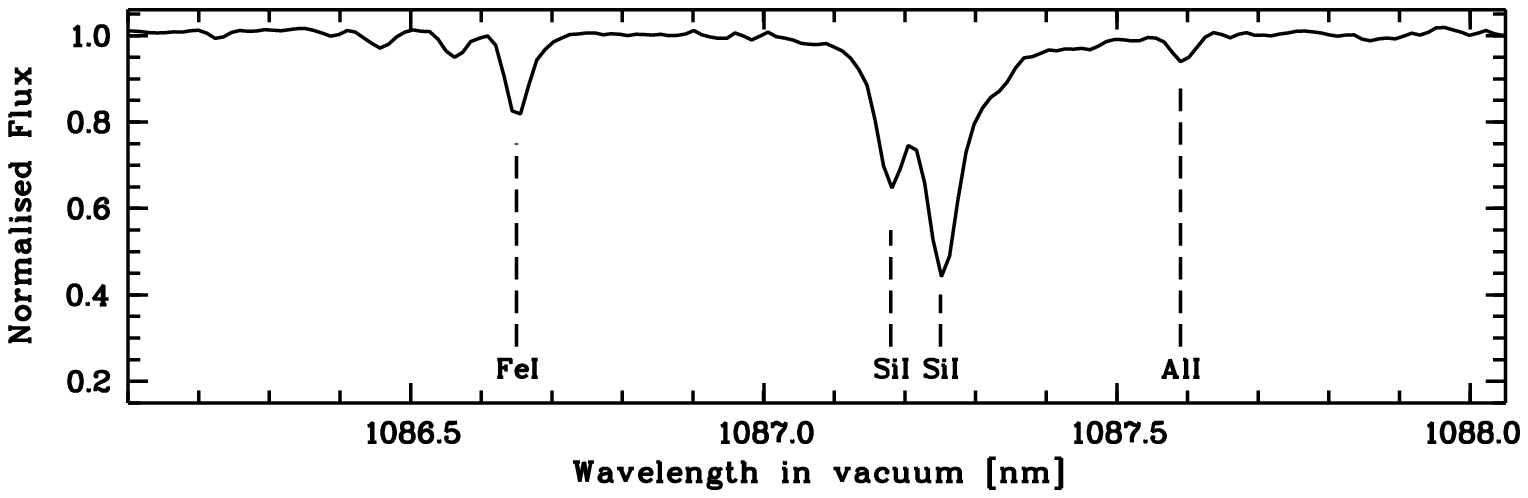}
      \caption{The \ion{Al}{i} line at 1087\,nm (solid black) in the case of the star HD\,24040.
}
         \label{al1087hd24040}
   \end{figure}

For Al, we investigated two lines at 1087.59 and 1676.79\,nm, which are formed close to LTE.
Our Al atomic model is described in detail in \citet{andrievsky08}.
This model atom consists of 76 levels of \ion{Al}{i} and 13 levels of
\ion{Al}{ii}. As shown by test calculations within that same grid star
parameters as for the other NLTE calculations, \ion{Al}{i} lines 1087 and
1676\,nm respond very poorly to NLTE effects. The effect on EW is
within $\pm 4$\%, depending on the parameters of the atmosphere.
For all stars but three, A(Al) is based only on the \ion{Al}{i} line at 1087.59\,nm;
this line is weak but is in a range that is not overly contaminated by telluric absorption (see Fig.\,\ref{al1087hd24040}).
For the three stars for which also the line at 1676.79\,nm is kept, the agreement in A(Al)
derived from the two lines is good.
For the 40 stars, we find $\langle{\rm [Al/Fe]}\rangle = +0.10\pm 0.06$,
with some stars with [Al/Fe] of about +0.2\,dex.
Only one star, HD\,136618, has a negative [Al/Fe]; this remains compatible within uncertainties with a solar ratio.

%-------------------------------------------------------------
%-------------------------------------------------------------

Five lines of \ion{K}{i} have been analysed, at 1177.28, 1177.50, 1243.56, 1252.55, and 1516.72\,nm.
The model atom of K was taken from \citet{andrievsky10} and
consisted of 20 levels of \ion{K}{i} and the ground level of \ion{K}{ii}. In addition,
another 15 levels of \ion{K}{i} and seven levels of \ion{K}{ii} were used to ensure
particle number conservation. The total number of bound-bound
radiative transitions taken into account was 62 \citep[see][for further details]{andrievsky10}.  Infrared lines of potassium showed a strong susceptibility to the influence
of NLTE effects. Therefore, due to deviations from the LTE, the EW of the lines used increased from $10$\% to $32$\%.
 We were able to derive A(K) for all but two stars, and $\langle{\rm [K/Fe]}\rangle = +0.16\pm 0.06$ for the 38 stars for which K has been derived.
All the lines are sensitive to NLTE, and therefore the above-solar [K/Fe] value can easily be attributed to our LTE assumption.

A(Ca) from \ion{Ca}{i} and \ion{Ca}{ii} lines is in good agreement 
($\langle{\rm A(\ion{Ca}{i})-A(\ion{Ca}{ii})}\rangle = -0.04\pm 0.08$)
with a slightly higher value of A(Ca) when derived from the \ion{Ca}{ii} lines, which can be explain
by the fact that the \ion{Ca}{i} lines we selected to derive the Ca abundance 
are close to LTE while some of the \ion{Ca}{ii} lines are sensitive to NLTE and the A(Ca) derived
in LTE overestimates the abundance.
However, the A(Ca) abundance we derived for the sample stars is slightly high:
from \ion{Ca}{i} lines, $\langle{[\rm Ca/Fe]}\rangle = +0.12\pm 0.05$.
In particular for four stars, HD\,97645, HD\,99492, HD\,147231 and HD\,190360, ${[\rm Ca/Fe]} > 0.2$.
The latter star of the four is overabundant in several elements (see below).

The Ti abundances from \ion{Ti}{i} and \ion{Ti}{ii} lines
($\langle{\rm A(\ion{Ti}{i})-A(\ion{Ti}{ii})}\rangle = +0.10\pm 0.08$) are well in agreement
within uncertainties, and the [Ti/Fe] for the stars analysed 
from \ion{Ti}{i} lines is $\langle{[\rm Ti/Fe]}\rangle = +0.09\pm 0.05$.
HD\,116321 is rich in Ti.

All three \ion{Sr}{ii} lines already investigated in  \citet{phosphoro2016} are affected by NLTE. 
The high [Sr/Fe] we derive for the 37 stars for which we derive A(Sr) ($\langle[\rm {Sr/Fe}]\rangle = +0.42\pm 0.12$)
is due to NLTE effects.

%-------------------------------------- 
   \begin{figure}
   \centering
   \includegraphics[width=\hsize,clip=true]{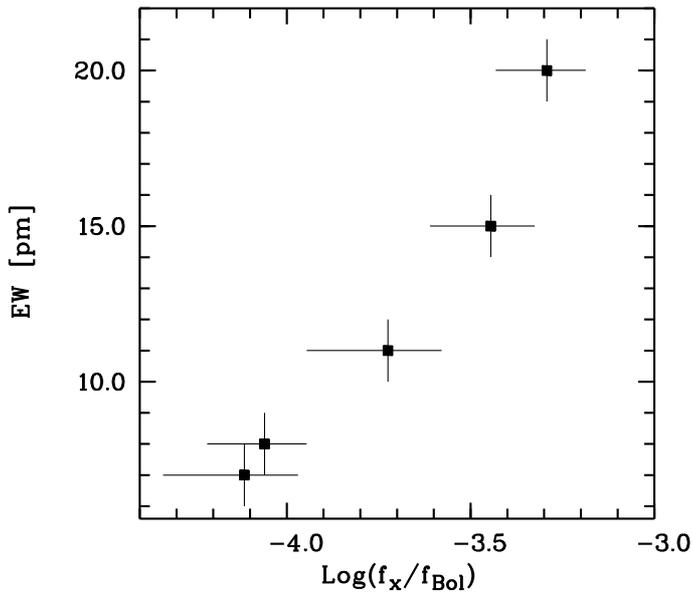}
   \caption{The EW of the \ion{He}{i} line vs. the logarithm of the
ratio of X and bolometric flux.}
              \label{ewfxf}%
    \end{figure}
%-------------------------------------- 

One star, HD\,190360, has [P/Fe], [C/Fe], [S/Fe], and [Ca/Fe] ratios that are  clearly high with respect to the solar pattern.
This could be due to an underestimated effective temperature, HD\,190360 being the only star for which the temperature is derived
from a single calibration.
Increasing the temperature by 180\,K to match for example the \Teff\ of \citet{mishenina04} of 5606\,K, 
the star no longer shows overabundance in C, P, and S, but still shows an overabundance  in Ca.
Also \citet{bensby03}, with an effective temperature of 5490\,K and a surface gravity of 4.23 
(values very similar
to the ones we adopted) found an overabundance with respect to Fe of several elements.
In \citet{bensby14} however, with a $\Teff = 5572$\,K, this overabundance is no longer present.
It is interesting to note that a linear regression of the mean \Teff\
over the $B-V$ \Teff\ for the 37 stars for which five temperature determinations are available 
predicts \Teff = 5593\,K very close to the values of  
\citet{mishenina04} and \citet{bensby14}.

%%%%%%%%%%%%%%%%%%%%%%%%%%%%%%%%%%%%%%%%%%%%%%%%%%%%%%%%%%%%%%%%%%%%%%%%
\subsection{Helium}

The triplet line of neutral He arising from the high-excitation (19.72\,eV) metastable level 
$\rm 2s^3S_1$ is observed in absorption in the Sun and in solar-type stars. It is of too high an excitation to be formed
in the photosphere and therefore must be formed in the chromosphere and very likely excited by radiation from the corona. 
In spatially resolved spectra of the Sun it can be observed in absorption in active
regions (plages). Following the activity cycle of the Sun, the EW measured in the
flux-integrated spectrum varies depending on the number of active regions. \citet{andretta} 
suggested to use this line together with \ion{He}{i} 587.6\,nm to estimate the fraction
of the stellar surface covered by active regions.
In a pioneering work, \citet{Zirin} obtained IR spectra at the coud{\'e} focus of the Palomar five-metre telescope
using image cathode tubes and photographic plates and films as detectors for 455 stars. 
The EW of the \ion{He}{i} 1083.0\,nm line is well correlated with the soft X-ray luminosity and 
in spite of the technical advances even today, the ``observational data on the $\lambda$10830 triplet line
remains sparse'' \citep[{\it verbatim} from][]{Smith}.
This line has not been studied in great detail because it requires high-resolution observations. 
The large spectral coverage provided by GIANO ensures that the spectral 
range is always observed, and it lies in a region
that is 
contaminated by a few 
%\LEt{please note the difference between "few" and "a few" and verify that this correction retains the intended meaning.}
telluric lines that sometimes affect the He line. 
We inspected all the spectra of our sample and were able to detect this line for 16 stars. 
When the line was detected we measured its EW (Table\,\ref{tabhe}).
The typical uncertainty is 1\,pm.
For five stars in our sample we found X-ray measurements in the All-Sky Survey Faint source Catalogue
\citep{Voges}. From these  we derived $\log(f_x/f_{bol}$),  also provided in Table\,\ref{tabhe}.
The apparent bolometric magnitude was derived from the  Gaia $G$ magnitude 
and the bolometric corrections of \citet{Bonifacio18}.
The bolometric flux was then derived as recommended in IAU resolution 2015 
B2\footnote{\url{https://www.iau.org/static/resolutions/IAU2015_English.pdf}}  equation\,(4).
These values are plotted in Fig.\ref{ewfxf} against the EW of the \ion{He}{i} 1083.0\,nm line
and a very good correlation can be seen between the two quantities.
GIANO observations offer a good opportunity to monitor the activity cycles of solar-type stars
from the observation of this line. 
This line may also be used to monitor the activity of stars with candidate
planets found in radial-velocity searches. 
The monitoring of the \ion{He}{i} 1083.0\,nm line should allow us to derive the activity
cycle of the star. This may help decipher whether  radial velocity
variations are due to planets or stellar activity. A well known example is
HD\,166435 \citep{Queloz}.

%%%%%%%%%%%%%%%%%%%%%%%%%%%% Table %%%%%%%%%%%%%%%%%%%%%%%%%%%%%%%%%%%%%
\begin{table}
\caption{Equivalent widths of the \ion{He}{i} 1083\,nm line.}
\label{tabhe}
\renewcommand{\tabcolsep}{3pt}
\tabskip=1pt
\begin{center}
\begin{tabular}{lrrr}
\hline\hline
\hline \noalign{\smallskip}
  Star  &     EW [pm] & $\log_{10}\left({\rm f_{\rm X}/f_{\rm Bol}}\right)$ & d$^a$ \\
\hline \noalign{\smallskip}
    HD\,20670  & $\sim 2$ \\
    HD\,24040  & $<5$ \\
    HD\,28005  & $<10$ \\ 
    HD\,32673  & $<10$ \\ 
    HD\,34445  & $2$ \\  
    HD\,34575  & $2$ \\  
    HD\,44420  & $<5$ \\
    HD\,56303  & $<10$ \\ 
    HD\,67346  & $<5$ \\
    HD\,69056  & $<10$ \\ 
    HD\,69809  & $<10$ \\ 
    HD\,69960  & $<10$ \\ 
    HD\,73226  & $<10$ \\ 
    HD\,73933  & $11$ \\
    HD\,76909  & $<5$ \\
    HD\,77519  & $<5$ \\
    HD\,82943  & $<10$ \\ 
    HD\,85301  & $20$ & $-3.29$ & 15.31\\
    HD\,87359  & $6$ \\ 
    HD\,87836  & $3$ \\ 
    HD\,90681  & $8$ \\
    HD\,90722  & $<10$ \\ 
    HD\,92788  & $4$ \\  
    HD\,97645  & $<5$ \\
    HD\,98618  & $<10$ \\ 
    HD\,98736  & $7$ \\  
    HD\,99491  & $7$ & $-4.12$ & 26.17 \\
    HD\,99492  & $11$ & $-3.72$ & 50.60\\
    HD\,100069 & $<5$ \\
    HD\,105631 & $15$ & $-3.45$ & 1.96\\
    HD\,106116 & $<5$ \\
    HD\,106156 & $12$ \\
    HD\,108942 & $3$ \\ 
    HD\,114174 & $<5$ \\
    HD\,116321 & $<5$ \\
    HD\,136618 & $<5$ \\
    HD\,145675 & $6$ \\
    HD\,147231 & $<5$ \\
    HD\,159222 & $8$ & $-4.06$ & 18.85\\
    HD\,161797 & $<5$ \\
\noalign{\smallskip}
\hline
\multicolumn{4}{l}{d$^a$ angular distance $''$ between}\\
\multicolumn{4}{l}{the X-ray source and the star}\\
\hline
\end{tabular}
\end{center}
\end{table}
%%%%%%%%%%%%%%%%%%%%%%%%%%%%%%%%%%%%%%%%%%%%%%%%%%%%%%%%%%%%%%%%%%%%%%%%

%%%%%%%%%%%%%%%%%%%%%%%%%%%%%%%%%%%%%%%%%%%%%%%%%%%%%%%%%%%%%%%%%%%%%%%%
\section{Conclusions}

In this work, we analysed 40 stars for which we provided detailed chemical abundances of 14 elements.
These stars belong to the Galactic disc and the abundances we derive are consistent with
the chemical behaviour expected for Galactic disc stars; they are also in line with previous investigations
\citep[e.g. see][]{mikolaitis14,smiljanic14,smiljanic16,kordopatis15,duffau17}.
For some of the stars investigated here there are no detailed chemical analyses in the literature.

We were also able to add 40 values for the P abundance in Galactic stars, which is still a rather poorly studied element.
We also introduced a \ion{P}{i} line of Mult.\,4, the abundance of which is consistent with, though slightly higher than, the
P abundance derived from the \ion{P}{i} lines of Mult.\,1, used in previous investigations
\citep{phosphoro2011,phosphoro2016}.

For a subsample of nine stars, we compared the abundances derived from the IR GIANO spectra and SOPHIE optical spectra.
For eight chemical abundances common to both surveys agreement was found to be very good.
Nevertheless, the line-to-line scatter in the IR spectra is larger than in the optical spectra, probably due to a better
knowledge on atomic data in the latter range and this is probably related to the longer history of use of optical ranges.

Both the good agreement on the abundances derived from IR and optical spectra and the good agreement with the
literature are indicators that the abundances we derive from GIANO spectra are reliable.
For the star HD\,98736, we analysed only four orders of GIANO (47-50) which correspond to the high-resolution
mode of the H-band for MOONS. In this case we can derive abundances for seven elements (C, Si, S, Ti, Cr, Fe and Ni). 
We analysed this range also for the star HD\,24040, the spectral range shown in Fig.\,\ref{o73},
and we derive the abundance for C, Si, S, K, Ti, Cr, Fe, Ni.
Clearly, by using only four orders of the GIANO spectra, the uncertainties in the abundances are much larger
because the number of lines available is much lower than when analysing the complete GIANO range;
for the majority of the elements abundances are based on a single line.
On top of that, because of the lower resolution that MOONS will have with respect to GIANO, some of these
elements will be difficult to be derived by MOONS, in particular Ti, Cr, and Ni.

For several stars in the sample we were able to detect the \ion{He}{i} line at 1083.0\,nm and confirm that its 
EW is strongly correlated with the logarithmic
ratio of X-ray flux to bolometric flux and, hence,
stellar activity.
Since He measurements in dwarf stars are still limited in number, GIANO provides a great opportunity
in regards to this topic.

GIANO provides us with a great chance to investigate Galactic stars and to derive their detailed chemical abundances.
From GIANO spectra we could derive the abundances of some elements (C, P, K, Sr) that could not be derived from the
SOPHIE optical spectra.
GIANO's high resolution and large spectral range are fundamental to deriving trustable abundances for several elements.
GIANO is a single-object spectrograph meaning that stars can only be observed one at a time, but in the next years its archive will fill and the number
of stars available will be larger and larger.

\begin{acknowledgements}
We are grateful to the referee, Mathieu Van der Swaelmen, for his
suggestions and comments which made the paper more readable and understandable.
EC and PB have been supported by the Programme National de Physique Stellaire
of the Institut National des Sciences de l'Univers of CNRS.
SD  acknowledges  support from  Comit\'e Mixto  ESO-Gobierno  de  Chile. 
\end{acknowledgements}

% WARNING
%-------------------------------------------------------------------
% Please note that we have included the references to the file aa.dem in
% order to compile it, but we ask you to:
%
% - use BibTeX with the regular commands:
%   \bibliographystyle{aa} % style aa.bst
%   \bibliography{Yourfile} % your references Yourfile.bib
%
% - join the .bib files when you upload your source files
%-------------------------------------------------------------------

\begin{appendix}
\section{Stellar parameters and chemical abundances}

%%%%%%%%%%%%%%%%%%%%%%%%%%%%% Table %%%%%%%%%%%%%%%%%%%%%%%%%%%%%%%%%%%%%
\begin{table*}
\caption{ Stellar parameters of our complete sample of stars. The uncertainty in gravity is smaller than 0.1\,dex for all stars.}
\label{fevalues}
\renewcommand{\tabcolsep}{3pt}
\tabskip=1pt
\begin{center}
\begin{tabular}{lllrllll}
\hline\hline
\hline \noalign{\smallskip}
{Star}   & {\Teff} & \logg  & {[Fe/H]} & A(FeI) & N lines & A(FeII) & N line \\
         &     K   &  [cgs] &       & \\
\hline \noalign{\smallskip}
 HD\,20670   & $5688\pm 138$ & 3.65 & $ 0.10$ &  $7.62\pm 0.16$  & 76  &  $7.63\pm 0.09$  &  4 \\
 HD\,24040   & $5809\pm 136$ & 4.12 & $ 0.09$ &  $7.61\pm 0.17$  & 81  &  $7.60\pm 0.07$  &  4 \\
 HD\,28005   & $5802\pm 144$ & 4.18 & $ 0.21$ &  $7.73\pm 0.16$  & 71  &  $7.73\pm 0.09$  &  4 \\
 HD\,32673   & $5752\pm 135$ & 3.53 & $ 0.07$ &  $7.59\pm 0.15$  & 73  &  $7.53\pm 0.09$  &  4 \\
 HD\,34445   & $5803\pm 112$ & 4.06 & $-0.03$ &  $7.49\pm 0.15$  & 76  &  $7.51\pm 0.11$  &  3 \\
 HD\,34575   & $5582\pm 134$ & 4.22 & $ 0.18$ &  $7.70\pm 0.17$  & 60  &  $7.74\pm 0.06$  &  3 \\
 HD\,44420   & $5777\pm 151$ & 4.23 & $ 0.19$ &  $7.71\pm 0.16$  & 65  &  $7.68\pm 0.09$  &  4 \\
 HD\,56303   & $5941\pm 125$ & 4.21 & $ 0.05$ &  $7.57\pm 0.15$  & 90  &  $7.59\pm 0.09$  &  4 \\
 HD\,67346   & $5953\pm 140$ & 3.78 & $ 0.14$ &  $7.66\pm 0.16$  & 80  &  $7.66\pm 0.11$  &  4 \\
 HD\,69056   & $5637\pm 138$ & 4.28 & $ 0.05$ &  $7.57\pm 0.16$  & 78  &  $7.55\pm 0.12$  &  3 \\
 HD\,69809   & $5842\pm 147$ & 4.16 & $ 0.17$ &  $7.69\pm 0.14$  & 75  &  $7.68\pm 0.14$  &  4 \\
 HD\,69960   & $5655\pm 147$ & 3.99 & $ 0.22$ &  $7.74\pm 0.17$  & 62  &  $7.67\pm 0.06$  &  3 \\
 HD\,73226   & $5886\pm 148$ & 4.21 & $ 0.06$ &  $7.58\pm 0.16$  & 81  &  $7.54\pm 0.14$  &  3 \\
 HD\,73933   & $6143\pm 134$ & 4.24 & $ 0.06$ &  $7.58\pm 0.17$  & 83  &  $7.49\pm 0.14$  &  3 \\
 HD\,76909   & $5655\pm 159$ & 4.17 & $ 0.24$ &  $7.76\pm 0.17$  & 60  &  $7.82\pm 0.06$  &  2 \\
 HD\,77519   & $6140\pm 190$ & 3.85 & $ 0.14$ &  $7.66\pm 0.17$  & 94  &  $7.66\pm 0.12$  &  4 \\
 HD\,82943   & $5917\pm 284$ & 4.23 & $ 0.13$ &  $7.65\pm 0.17$  & 84  &  $7.70\pm 0.05$  &  4 \\
 HD\,85301   & $5640\pm 128$ & 4.44 & $ 0.05$ &  $7.57\pm 0.19$  & 70  &  $7.61\pm 0.02$  &  2 \\
 HD\,87359   & $5645\pm 140$ & 4.40 & $-0.07$ &  $7.45\pm 0.17$  & 78  &  $7.50\pm 0.05$  &  4 \\
 HD\,87836   & $5684\pm 158$ & 4.05 & $ 0.16$ &  $7.68\pm 0.15$  & 58  &  $7.63\pm 0.17$  &  4 \\
 HD\,90681   & $5950\pm 167$ & 4.26 & $ 0.16$ &  $7.68\pm 0.16$  & 78  &  $7.68\pm 0.09$  &  4 \\
 HD\,90722   & $5677\pm 141$ & 4.12 & $ 0.22$ &  $7.74\pm 0.16$  & 57  &  $7.70        $  &  1 \\
 HD\,92788   & $5694\pm 196$ & 4.26 & $ 0.13$ &  $7.65\pm 0.17$  & 68  &  $7.67\pm 0.16$  &  4 \\
 HD\,97645   & $6127\pm 125$ & 4.10 & $ 0.14$ &  $7.66\pm 0.18$  & 81  &  $7.62\pm 0.09$  &  4 \\
 HD\,98618   & $5727\pm 138$ & 4.27 & $-0.11$ &  $7.41\pm 0.17$  & 89  &  $7.45\pm 0.05$  &  3 \\
 HD\,98736   & $5276\pm 100$ & 4.40 & $ 0.29$ &  $7.81\pm 0.19$  & 49  &  $7.90\pm 0.13$  &  3 \\
 HD\,99491   & $5537\pm 178$ & 4.40 & $ 0.25$ &  $7.77\pm 0.18$  & 58  &  $7.76\pm 0.14$  &  2 \\
 HD\,99492   & $5006\pm 235$ & 4.56 & $ 0.20$ &  $7.72\pm 0.23$  & 52  &  $7.75\pm 0.12$  &  2 \\
 HD\,100069  & $5796\pm 129$ & 3.74 & $-0.03$ &  $7.49\pm 0.18$  & 79  &  $7.45\pm 0.15$  &  3 \\
 HD\,105631  & $5391\pm 104$ & 4.47 & $ 0.05$ &  $7.57\pm 0.18$  & 62  &  $7.61\pm 0.05$  &  2 \\
 HD\,106116  & $5665\pm 164$ & 4.30 & $ 0.03$ &  $7.55\pm 0.16$  & 70  &  $7.61\pm 0.11$  &  3 \\
 HD\,106156  & $5449\pm 175$ & 4.45 & $ 0.10$ &  $7.62\pm 0.17$  & 61  &  $7.67        $  &  1 \\
 HD\,108942  & $5882\pm 179$ & 4.27 & $ 0.20$ &  $7.72\pm 0.18$  & 74  &  $7.66\pm 0.10$  &  4 \\
 HD\,114174  & $5728\pm 193$ & 4.30 & $-0.06$ &  $7.46\pm 0.17$  & 74  &  $7.48\pm 0.08$  &  3 \\
 HD\,116321  & $6292\pm 188$ & 3.66 & $ 0.10$ &  $7.62\pm 0.20$  & 81  &  $7.56\pm 0.10$  &  3 \\
 HD\,136618  & $5805\pm 125$ & 3.56 & $ 0.14$ &  $7.66\pm 0.18$  & 73  &  $7.62\pm 0.15$  &  4 \\
 HD\,145675  & $5312\pm 149$ & 4.37 & $ 0.33$ &  $7.85\pm 0.18$  & 46  &  $7.92\pm 0.05$  &  2 \\
 HD\,147231  & $5594\pm 132$ & 4.30 & $-0.14$ &  $7.38\pm 0.17$  & 78  &  $7.48\pm 0.09$  &  3 \\
 HD\,159222  & $5815\pm 125$ & 4.30 & $ 0.00$ &  $7.52\pm 0.16$  & 76  &  $7.57\pm 0.11$  &  4 \\
 HD\,190360  & $5424\pm 250$ & 4.21 & $-0.01$ &  $7.51\pm 0.16$  & 57  &  $7.59\pm 0.06$  &  3 \\
\noalign{\smallskip}
\hline
\end{tabular}
\end{center}
\end{table*}
%%%%%%%%%%%%%%%%%%%%%%%%%%%%%%%%%%%%%%%%%%%%%%%%%%%%%%%%%%%%%%%%%%%%%%%%

%%%%%%%%%%%%%%%%%%%%%%%%%%%% Table %%%%%%%%%%%%%%%%%%%%%%%%%%%%%%%%%%%%%
\begin{table*}
\caption{Abundances for C, Na, Mg and Al for our complete sample of stars.}
\label{abbocnaal}
\renewcommand{\tabcolsep}{3pt}
\tabskip=1pt
\begin{center}
\begin{tabular}{llllllllll}
\hline\hline
\hline \noalign{\smallskip}
{Star}   & A(C) & N lines  & A(Na) & N lines & A(Mg) & N lines & A(Al) & N lines \\
\hline \noalign{\smallskip}
 HD\,20670   & $8.50\pm 0.22$  & 22  &  $6.41\pm 0.07$  &  3  &  $7.69\pm 0.11$  & 10  &  $6.71        $  &  1 \\ 
 HD\,24040   & $8.48\pm 0.21$  & 22  &  $6.33\pm 0.07$  &  2  &  $7.68\pm 0.13$  & 10  &  $6.62        $  &  1 \\ 
 HD\,28005   & $8.68\pm 0.23$  & 19  &  $6.60        $  &  1  &  $7.81\pm 0.13$  &  8  &  $6.85        $  &  1 \\ 
 HD\,32673   & $8.44\pm 0.23$  & 22  &  $6.35\pm 0.06$  &  2  &  $7.69\pm 0.08$  &  8  &  $6.57        $  &  1 \\ 
 HD\,34445   & $8.44\pm 0.22$  & 24  &  $6.22        $  &  1  &  $7.56\pm 0.12$  &  8  &  $6.49        $  &  1 \\ 
 HD\,34575   & $8.64\pm 0.24$  & 23  &  $6.42\pm 0.11$  &  2  &  $7.82\pm 0.09$  &  8  &  $6.78        $  &  1 \\ 
 HD\,44420   & $8.65\pm 0.24$  & 21  &  $6.42\pm 0.19$  &  2  &  $7.80\pm 0.10$  &  9  &  $6.87        $  &  1 \\ 
 HD\,56303   & $8.44\pm 0.19$  & 25  &  $6.19\pm 0.03$  &  2  &  $7.65\pm 0.10$  &  9  &  $6.56        $  &  1 \\ 
 HD\,67346   & $8.61\pm 0.20$  & 21  &  $6.52\pm 0.09$  &  2  &  $7.70\pm 0.14$  & 10  &  $6.63        $  &  1 \\ 
 HD\,69056   & $8.50\pm 0.22$  & 25  &  $6.22\pm 0.15$  &  3  &  $7.69\pm 0.12$  & 10  &  $6.69        $  &  1 \\ 
 HD\,69809   & $8.65\pm 0.22$  & 19  &  $6.44        $  &  1  &  $7.73\pm 0.13$  & 11  &  $6.85        $  &  1 \\ 
 HD\,69960   & $8.64\pm 0.22$  & 21  &  $6.29\pm 0.15$  &  2  &  $7.66\pm 0.29$  &  8  &  $6.83        $  &  1 \\ 
 HD\,73226   & $8.48\pm 0.21$  & 24  &  $6.24\pm 0.23$  &  3  &  $7.60\pm 0.14$  &  7  &  $6.68        $  &  1 \\ 
 HD\,73933   & $8.45\pm 0.21$  & 24  &  $6.17\pm 0.00$  &  2  &  $7.59\pm 0.14$  & 12  &  $6.59\pm 0.11$  &  2 \\ 
 HD\,76909   & $8.73\pm 0.20$  & 19  &  $6.51        $  &  1  &  $7.82\pm 0.12$  &  8  &  $6.81        $  &  1 \\ 
 HD\,77519   & $8.57\pm 0.23$  & 20  &  $6.30\pm 0.19$  &  3  &  $7.68\pm 0.14$  & 10  &  $6.68\pm 0.12$  &  2 \\ 
 HD\,82943   & $8.57\pm 0.21$  & 21  &  $6.40\pm 0.07$  &  2  &  $7.67\pm 0.14$  &  9  &  $6.72        $  &  1 \\ 
 HD\,85301   & $8.45\pm 0.21$  & 30  &  $6.08\pm 0.03$  &  2  &  $7.53\pm 0.18$  &  9  &  $6.59        $  &  1 \\ 
 HD\,87359   & $8.36\pm 0.19$  & 34  &  $6.13\pm 0.06$  &  2  &  $7.49\pm 0.11$  & 10  &  $6.40        $  &  1 \\ 
 HD\,87836   & $8.66\pm 0.26$  & 23  &  $6.60        $  &  1  &  $7.73\pm 0.14$  &  8  &  $6.77        $  &  1 \\ 
 HD\,90681   & $8.55\pm 0.21$  & 22  &  $6.34\pm 0.16$  &  3  &  $7.73\pm 0.14$  & 10  &  $6.68        $  &  1 \\ 
 HD\,90722   & $8.73\pm 0.21$  & 21  &  $6.50        $  &  1  &  $7.80\pm 0.09$  &  6  &  $6.78        $  &  1 \\ 
 HD\,92788   & $8.63\pm 0.17$  & 21  &  $6.46\pm 0.15$  &  2  &  $7.71\pm 0.17$  & 11  &  $6.79        $  &  1 \\ 
 HD\,97645   & $8.48\pm 0.22$  & 21  &  $6.27\pm 0.00$  &  2  &  $7.66\pm 0.11$  & 11  &  $6.75\pm 0.02$  &  2 \\ 
 HD\,98618   & $8.41\pm 0.21$  & 27  &  $6.13\pm 0.22$  &  3  &  $7.49\pm 0.11$  & 11  &  $6.42        $  &  1 \\ 
 HD\,98736   & $8.84\pm 0.24$  & 27  &  $6.72        $  &  1  &  $7.81\pm 0.15$  &  6  &  $6.89        $  &  1 \\ 
 HD\,99491   & $8.64\pm 0.17$  & 26  &  $6.53\pm 0.18$  &  2  &  $7.77\pm 0.18$  &  8  &  $6.86        $  &  1 \\ 
 HD\,99492   & $8.71\pm 0.35$  & 28  &  $6.62\pm 0.14$  &  2  &  $7.68\pm 0.21$  &  5  &  $6.90        $  &  1 \\ 
 HD\,100069  & $8.45\pm 0.22$  & 22  &  $6.19\pm 0.08$  &  3  &  $7.37\pm 0.48$  &  7  &  $6.58        $  &  1 \\ 
 HD\,105631  & $8.49\pm 0.23$  & 32  &  $6.27\pm 0.02$  &  2  &  $7.57\pm 0.15$  &  9  &  $6.54        $  &  1 \\ 
 HD\,106116  & $8.50\pm 0.16$  & 25  &  $6.27\pm 0.11$  &  3  &  $7.43\pm 0.51$  &  7  &  $6.66        $  &  1 \\ 
 HD\,106156  & $8.60\pm 0.16$  & 26  &  $6.32\pm 0.07$  &  3  &  $7.59\pm 0.15$  &  8  &  $6.64        $  &  1 \\ 
 HD\,108942  & $8.57\pm 0.23$  & 22  &  $6.36        $  &  1  &  $7.73\pm 0.14$  & 10  &  $6.75        $  &  1 \\ 
 HD\,114174  & $8.47\pm 0.20$  & 25  &  $6.23\pm 0.12$  &  2  &  $7.38\pm 0.42$  &  7  &  $6.50        $  &  1 \\ 
 HD\,116321  & $8.55\pm 0.24$  & 17  &  $            $  &  0  &  $7.67\pm 0.12$  &  8  &  $6.71        $  &  1 \\ 
 HD\,136618  & $8.59\pm 0.20$  & 18  &  $6.45        $  &  1  &  $7.76\pm 0.14$  &  9  &  $6.53        $  &  1 \\ 
 HD\,145675  & $8.84\pm 0.22$  & 27  &  $6.76        $  &  1  &  $7.90\pm 0.11$  &  4  &  $6.94        $  &  1 \\ 
 HD\,147231  & $8.47\pm 0.25$  & 29  &  $6.06\pm 0.17$  &  3  &  $7.56\pm 0.11$  &  9  &  $6.40        $  &  1 \\ 
 HD\,159222  & $8.47\pm 0.22$  & 22  &  $6.16\pm 0.07$  &  3  &  $7.54\pm 0.18$  &  9  &  $6.53        $  &  1 \\ 
 HD\,190360  & $8.84\pm 0.16$  & 21  &  $6.26\pm 0.08$  &  2  &  $7.67\pm 0.10$  &  4  &  $6.64        $  &  1 \\ 
\noalign{\smallskip}
\hline
\end{tabular}
\end{center}
\end{table*}
%%%%%%%%%%%%%%%%%%%%%%%%%%%%%%%%%%%%%%%%%%%%%%%%%%%%%%%%%%%%%%%%%%%%%%%%

%%%%%%%%%%%%%%%%%%%%%%%%%%%% Table %%%%%%%%%%%%%%%%%%%%%%%%%%%%%%%%%%%%%
\begin{table*}
\caption{Continued: abundances for Si, P, S and K.}
\label{abbosips}
\renewcommand{\tabcolsep}{3pt}
\tabskip=1pt
\begin{center}
\begin{tabular}{llllllllll}
\hline\hline
\hline \noalign{\smallskip}
{Star}   & A(Si) & N lines  & A(P) & N lines & A(S) & N lines & A(K) & N lines \\
\hline \noalign{\smallskip}
 HD\,20670   & $7.61\pm 0.24$  & 28  &  $5.65\pm 0.12$  &  2  &  $7.27\pm 0.15$  &  8  & $5.39\pm 0.08$  &  3  \\ 
 HD\,24040   & $7.61\pm 0.23$  & 28  &  $5.60\pm 0.13$  &  3  &  $7.25\pm 0.17$  &  8  & $5.37\pm 0.22$  &  5  \\ 
 HD\,28005   & $7.70\pm 0.23$  & 24  &  $5.79\pm 0.10$  &  3  &  $7.35\pm 0.17$  &  8  & $5.57\pm 0.08$  &  2  \\ 
 HD\,32673   & $7.56\pm 0.22$  & 26  &  $5.49\pm 0.17$  &  3  &  $7.28\pm 0.21$  &  7  & $5.37\pm 0.05$  &  3  \\ 
 HD\,34445   & $7.54\pm 0.20$  & 24  &  $5.52\pm 0.04$  &  3  &  $7.24\pm 0.12$  &  9  & $5.26\pm 0.15$  &  4  \\ 
 HD\,34575   & $7.65\pm 0.23$  & 23  &  $5.74\pm 0.10$  &  3  &  $7.42\pm 0.17$  &  8  & $5.43\pm 0.13$  &  3  \\ 
 HD\,44420   & $7.71\pm 0.23$  & 27  &  $5.79\pm 0.13$  &  3  &  $7.47\pm 0.18$  & 10  & $5.57\pm 0.02$  &  2  \\ 
 HD\,56303   & $7.56\pm 0.23$  & 28  &  $5.54\pm 0.05$  &  3  &  $7.21\pm 0.13$  & 11  & $5.23\pm 0.15$  &  4  \\ 
 HD\,67346   & $7.70\pm 0.25$  & 26  &  $5.72\pm 0.05$  &  3  &  $7.46\pm 0.18$  &  6  & $5.40\pm 0.13$  &  4  \\ 
 HD\,69056   & $7.58\pm 0.23$  & 26  &  $5.51\pm 0.06$  &  3  &  $7.16\pm 0.17$  & 10  & $5.34\pm 0.13$  &  4  \\ 
 HD\,69809   & $7.70\pm 0.24$  & 24  &  $5.77\pm 0.09$  &  3  &  $7.34\pm 0.08$  & 10  & $5.51\pm 0.10$  &  3  \\ 
 HD\,69960   & $7.72\pm 0.27$  & 22  &  $5.73\pm 0.07$  &  3  &  $7.30\pm 0.18$  &  8  & $5.51\pm 0.02$  &  2  \\ 
 HD\,73226   & $7.63\pm 0.22$  & 27  &  $5.60\pm 0.03$  &  3  &  $7.28\pm 0.17$  &  9  & $5.28\pm 0.17$  &  3  \\ 
 HD\,73933   & $7.63\pm 0.21$  & 23  &  $5.45\pm 0.12$  &  2  &  $7.21\pm 0.16$  &  7  & $5.28\pm 0.13$  &  4  \\ 
 HD\,76909   & $7.74\pm 0.26$  & 19  &  $5.86\pm 0.08$  &  3  &  $7.39\pm 0.20$  &  9  & $5.58\pm 0.03$  &  2  \\ 
 HD\,77519   & $7.68\pm 0.26$  & 27  &  $5.75\pm 0.05$  &  3  &  $7.37\pm 0.15$  &  7  & $5.45\pm 0.23$  &  5  \\ 
 HD\,82943   & $7.67\pm 0.24$  & 26  &  $5.67\pm 0.08$  &  3  &  $7.33\pm 0.18$  &  9  & $5.37\pm 0.09$  &  4  \\ 
 HD\,85301   & $7.59\pm 0.25$  & 24  &  $5.53\pm 0.08$  &  2  &  $7.29\pm 0.24$  &  8  & $5.17\pm 0.16$  &  3  \\ 
 HD\,87359   & $7.50\pm 0.23$  & 27  &  $5.46\pm 0.20$  &  3  &  $7.12\pm 0.16$  & 12  & $5.12\pm 0.15$  &  4  \\ 
 HD\,87836   & $7.70\pm 0.25$  & 25  &  $5.70\pm 0.15$  &  3  &  $7.32\pm 0.17$  &  8  & $5.50\pm 0.05$  &  2  \\ 
 HD\,90681   & $7.70\pm 0.25$  & 27  &  $5.65\pm 0.16$  &  3  &  $7.35\pm 0.17$  &  7  & $5.38\pm 0.14$  &  4  \\ 
 HD\,90722   & $7.69\pm 0.26$  & 24  &  $5.75\pm 0.17$  &  3  &  $7.41\pm 0.12$  &  8  & $5.41\pm 0.17$  &  2  \\ 
 HD\,92788   & $7.69\pm 0.25$  & 25  &  $5.74\pm 0.10$  &  3  &  $7.36\pm 0.17$  &  9  & $5.37\pm 0.13$  &  3  \\ 
 HD\,97645   & $7.72\pm 0.23$  & 26  &  $5.61\pm 0.10$  &  3  &  $7.28\pm 0.19$  &  9  & $5.34\pm 0.15$  &  4  \\ 
 HD\,98618   & $7.49\pm 0.21$  & 26  &  $5.49\pm 0.09$  &  3  &  $7.16\pm 0.22$  & 12  & $5.08\pm 0.09$  &  4  \\ 
 HD\,98736   & $7.85\pm 0.25$  & 21  &  $5.97\pm 0.25$  &  3  &  $7.52\pm 0.18$  &  7  & $5.62        $  &  1  \\ 
 HD\,99491   & $7.76\pm 0.24$  & 21  &  $5.65\pm 0.24$  &  3  &  $7.27\pm 0.23$  &  9  & $            $  &  0  \\ 
 HD\,99492   & $7.75\pm 0.27$  & 24  &  $5.72\pm 0.24$  &  2  &  $7.15\pm 0.31$  &  6  & $            $  &  0  \\ 
 HD\,100069  & $7.53\pm 0.22$  & 27  &  $5.53\pm 0.09$  &  3  &  $7.18\pm 0.12$  &  9  & $5.30\pm 0.23$  &  5  \\ 
 HD\,105631  & $7.56\pm 0.29$  & 24  &  $5.64\pm 0.36$  &  2  &  $7.23\pm 0.17$  & 11  & $5.22\pm 0.19$  &  3  \\ 
 HD\,106116  & $7.60\pm 0.24$  & 24  &  $5.64\pm 0.06$  &  3  &  $7.24\pm 0.16$  & 10  & $5.30        $  &  1  \\ 
 HD\,106156  & $7.64\pm 0.25$  & 23  &  $5.68\pm 0.10$  &  2  &  $7.33\pm 0.16$  &  7  & $5.34\pm 0.17$  &  3  \\ 
 HD\,108942  & $7.68\pm 0.26$  & 24  &  $5.76\pm 0.18$  &  3  &  $7.24\pm 0.16$  &  9  & $5.42\pm 0.19$  &  3  \\ 
 HD\,114174  & $7.52\pm 0.24$  & 23  &  $5.56\pm 0.03$  &  3  &  $7.11\pm 0.19$  & 10  & $5.28\pm 0.02$  &  3  \\ 
 HD\,116321  & $7.69\pm 0.22$  & 24  &  $5.71\pm 0.01$  &  2  &  $7.37\pm 0.21$  &  8  & $5.41\pm 0.18$  &  5  \\ 
 HD\,136618  & $7.67\pm 0.26$  & 20  &  $5.69\pm 0.06$  &  2  &  $7.24\pm 0.22$  &  8  & $5.41\pm 0.12$  &  3  \\ 
 HD\,145675  & $7.91\pm 0.30$  & 16  &  $5.91\pm 0.22$  &  3  &  $7.49\pm 0.17$  &  9  & $5.69        $  &  1  \\ 
 HD\,147231  & $7.49\pm 0.24$  & 22  &  $5.52\pm 0.09$  &  3  &  $7.12\pm 0.18$  & 10  & $5.23\pm 0.15$  &  4  \\ 
 HD\,159222  & $7.55\pm 0.21$  & 22  &  $5.53\pm 0.05$  &  3  &  $7.22\pm 0.16$  &  8  & $5.37\pm 0.10$  &  4  \\ 
 HD\,190360  & $7.67\pm 0.29$  & 20  &  $6.01\pm 0.38$  &  2  &  $7.52\pm 0.11$  &  4  & $5.29\pm 0.15$  &  3  \\ 
\noalign{\smallskip}
\hline
\end{tabular}
\end{center}
\end{table*}
%%%%%%%%%%%%%%%%%%%%%%%%%%%%%%%%%%%%%%%%%%%%%%%%%%%%%%%%%%%%%%%%%%%%%%%%

%%%%%%%%%%%%%%%%%%%%%%%%%%%% Table %%%%%%%%%%%%%%%%%%%%%%%%%%%%%%%%%%%%%
\begin{table*}
\caption{Continued: abundances for Ca and Ti.}
\label{abbokca}
\renewcommand{\tabcolsep}{3pt}
\tabskip=1pt
\begin{center}
\begin{tabular}{lllllllll}
\hline\hline
\hline \noalign{\smallskip}
{Star}   & A(CaI) & N lines  & A(CaII) & N lines & A(TiI) & N lines & A(TiII) & N lines \\
\hline \noalign{\smallskip}
 HD\,20670   & $ 6.55\pm 0.18$  &  4  & $ 6.55\pm 0.08$  &  4 & $5.07\pm 0.07$  &  5  & $ 4.90        $  &  1  \\
 HD\,24040   & $ 6.49\pm 0.12$  &  3  & $ 6.52\pm 0.06$  &  4 & $5.09\pm 0.10$  &  5  & $ 4.88        $  &  1  \\
 HD\,28005   & $ 6.60\pm 0.09$  &  3  & $ 6.62\pm 0.07$  &  3 & $5.18\pm 0.10$  &  5  & $ 4.99        $  &  1  \\
 HD\,32673   & $ 6.47\pm 0.14$  &  3  & $ 6.47\pm 0.16$  &  4 & $5.05\pm 0.19$  &  6  & $             $  &  0  \\
 HD\,34445   & $ 6.43\pm 0.17$  &  4  & $ 6.48\pm 0.07$  &  4 & $4.98\pm 0.09$  &  4  & $ 4.84        $  &  1  \\
 HD\,34575   & $ 6.63\pm 0.05$  &  3  & $ 6.60\pm 0.09$  &  4 & $5.13\pm 0.22$  &  6  & $ 5.05        $  &  1  \\
 HD\,44420   & $ 6.60\pm 0.05$  &  3  & $ 6.75\pm 0.05$  &  3 & $5.19\pm 0.06$  &  5  & $ 5.01        $  &  1  \\
 HD\,56303   & $ 6.53\pm 0.14$  &  4  & $ 6.57\pm 0.14$  &  4 & $4.99\pm 0.05$  &  5  & $ 4.94        $  &  1  \\
 HD\,67346   & $ 6.56\pm 0.15$  &  3  & $ 6.71\pm 0.09$  &  4 & $5.10\pm 0.07$  &  5  & $ 5.03        $  &  1  \\
 HD\,69056   & $ 6.50\pm 0.10$  &  3  & $ 6.51\pm 0.08$  &  4 & $5.07\pm 0.20$  &  6  & $ 4.97        $  &  1  \\
 HD\,69809   & $ 6.59\pm 0.07$  &  3  & $ 6.61\pm 0.11$  &  4 & $5.15\pm 0.04$  &  5  & $ 5.07        $  &  1  \\
 HD\,69960   & $ 6.62\pm 0.04$  &  3  & $ 6.59\pm 0.12$  &  3 & $5.15\pm 0.24$  &  6  & $ 4.98        $  &  1  \\
 HD\,73226   & $ 6.54\pm 0.10$  &  3  & $ 6.61\pm 0.14$  &  4 & $5.10\pm 0.17$  &  4  & $ 4.99        $  &  1  \\
 HD\,73933   & $ 6.54\pm 0.15$  &  4  & $ 6.60\pm 0.09$  &  3 & $5.07\pm 0.14$  &  4  & $ 5.00        $  &  1  \\
 HD\,76909   & $ 6.60\pm 0.10$  &  3  & $ 6.67\pm 0.01$  &  3 & $5.24\pm 0.06$  &  5  & $ 5.05        $  &  1  \\
 HD\,77519   & $ 6.63\pm 0.20$  &  4  & $ 6.76\pm 0.07$  &  4 & $5.11\pm 0.17$  &  4  & $ 4.99        $  &  1  \\
 HD\,82943   & $ 6.54\pm 0.07$  &  3  & $ 6.65\pm 0.11$  &  4 & $5.10\pm 0.06$  &  5  & $ 4.98        $  &  1  \\
 HD\,85301   & $ 6.55\pm 0.01$  &  3  & $ 6.55\pm 0.16$  &  4 & $5.02\pm 0.13$  &  5  & $ 5.01        $  &  1  \\
 HD\,87359   & $ 6.43\pm 0.07$  &  3  & $ 6.42\pm 0.09$  &  4 & $4.89\pm 0.19$  &  6  & $ 4.86        $  &  1  \\
 HD\,87836   & $ 6.61\pm 0.07$  &  3  & $ 6.64\pm 0.16$  &  3 & $5.07\pm 0.21$  &  5  & $ 4.95        $  &  1  \\
 HD\,90681   & $ 6.61\pm 0.12$  &  3  & $ 6.66\pm 0.03$  &  3 & $5.14\pm 0.10$  &  4  & $ 5.05        $  &  1  \\
 HD\,90722   & $ 6.62\pm 0.10$  &  3  & $ 6.67\pm 0.14$  &  3 & $5.15\pm 0.23$  &  5  & $ 5.18        $  &  1  \\
 HD\,92788   & $ 6.60\pm 0.07$  &  3  & $ 6.78\pm 0.38$  &  4 & $5.06\pm 0.21$  &  6  & $ 5.08        $  &  1  \\
 HD\,97645   & $ 6.69\pm 0.11$  &  4  & $ 6.62\pm 0.10$  &  4 & $5.16\pm 0.10$  &  3  & $ 5.09        $  &  1  \\
 HD\,98618   & $ 6.35\pm 0.09$  &  3  & $ 6.41\pm 0.11$  &  4 & $4.78\pm 0.14$  &  5  & $ 4.82        $  &  1  \\
 HD\,98736   & $ 6.68\pm 0.10$  &  3  & $ 6.73\pm 0.08$  &  3 & $5.24\pm 0.23$  &  4  & $ 5.05        $  &  1  \\
 HD\,99491   & $ 6.70\pm 0.09$  &  3  & $ 6.53\pm 0.20$  &  2 & $5.27\pm 0.19$  &  6  & $ 5.12        $  &  1  \\
 HD\,99492   & $ 6.74\pm 0.05$  &  3  & $ 6.74\pm 0.49$  &  5 & $5.20        $  &  1  & $ 5.11        $  &  1  \\
 HD\,100069  & $ 6.43\pm 0.13$  &  4  & $ 6.46\pm 0.13$  &  3 & $5.06        $  &  1  & $             $  &  0  \\
 HD\,105631  & $ 6.50\pm 0.12$  &  3  & $ 6.44\pm 0.04$  &  3 & $5.01\pm 0.22$  &  5  & $ 4.78        $  &  1  \\
 HD\,106116  & $ 6.52\pm 0.04$  &  3  & $ 6.50\pm 0.10$  &  4 & $4.94\pm 0.25$  &  5  & $ 4.88        $  &  1  \\
 HD\,106156  & $ 6.59\pm 0.08$  &  3  & $ 6.47\pm 0.25$  &  3 & $5.15\pm 0.20$  &  6  & $ 5.03        $  &  1  \\
 HD\,108942  & $ 6.60\pm 0.08$  &  3  & $ 6.51\pm 0.16$  &  4 & $5.19\pm 0.08$  &  4  & $ 4.98        $  &  1  \\
 HD\,114174  & $ 6.40\pm 0.04$  &  3  & $ 6.48\pm 0.11$  &  4 & $4.99\pm 0.05$  &  4  & $ 4.86        $  &  1  \\
 HD\,116321  & $ 6.52\pm 0.11$  &  4  & $ 6.68\pm 0.16$  &  4 & $5.26\pm 0.27$  &  3  & $ 5.33        $  &  1  \\
 HD\,136618  & $ 6.55\pm 0.15$  &  3  & $ 6.58\pm 0.11$  &  3 & $5.10\pm 0.08$  &  5  & $ 5.01        $  &  1  \\
 HD\,145675  & $ 6.73\pm 0.09$  &  3  & $ 6.83\pm 0.06$  &  3 & $5.37\pm 0.20$  &  5  & $             $  &  0  \\
 HD\,147231  & $ 6.45\pm 0.04$  &  3  & $ 6.34\pm 0.18$  &  5 & $4.92\pm 0.21$  &  5  & $ 4.82        $  &  1  \\
 HD\,159222  & $ 6.41\pm 0.10$  &  3  & $ 6.62\pm 0.05$  &  4 & $4.94\pm 0.02$  &  3  & $ 4.89        $  &  1  \\
 HD\,190360  & $ 6.59\pm 0.10$  &  3  & $ 6.71\pm 0.21$  &  4 & $5.01\pm 0.27$  &  5  & $ 5.08        $  &  1  \\
\noalign{\smallskip}
\hline
\end{tabular}
\end{center}
\end{table*}
%%%%%%%%%%%%%%%%%%%%%%%%%%%%%%%%%%%%%%%%%%%%%%%%%%%%%%%%%%%%%%%%%%%%%%%%

%%%%%%%%%%%%%%%%%%%%%%%%%%%% Table %%%%%%%%%%%%%%%%%%%%%%%%%%%%%%%%%%%%%
\begin{table*}
\caption{Continued: abundances for Cr, Ni and Sr.}
\label{abbocrnisr}
\renewcommand{\tabcolsep}{3pt}
\tabskip=1pt
\begin{center}
\begin{tabular}{lllllll}
\hline\hline
\hline \noalign{\smallskip}
{Star}   & A(Cr) & N lines  & A(Ni) & N lines & A(Sr) & N lines \\
\hline \noalign{\smallskip}
 HD\,20670   & $5.67\pm 0.13$  &  8  & $ 6.48\pm 0.17$  &  8  & $ 3.40        $  &  1 \\
 HD\,24040   & $5.65\pm 0.13$  &  9  & $ 6.50\pm 0.16$  &  9  & $ 3.29        $  &  1 \\
 HD\,28005   & $5.79\pm 0.09$  &  8  & $ 6.64\pm 0.19$  &  8  & $ 3.39        $  &  1 \\
 HD\,32673   & $5.64\pm 0.14$  &  8  & $ 6.46\pm 0.18$  &  9  & $ 3.34        $  &  1 \\
 HD\,34445   & $5.57\pm 0.11$  &  9  & $ 6.35\pm 0.21$  &  7  & $             $  &  0 \\
 HD\,34575   & $5.78\pm 0.10$  &  8  & $ 6.59\pm 0.20$  &  9  & $ 3.20        $  &  1 \\
 HD\,44420   & $5.76\pm 0.13$  &  8  & $ 6.64\pm 0.19$  &  8  & $ 3.45        $  &  1 \\
 HD\,56303   & $5.58\pm 0.11$  &  8  & $ 6.36\pm 0.20$  & 10  & $ 3.42        $  &  1 \\
 HD\,67346   & $5.70\pm 0.10$  &  6  & $ 6.54\pm 0.26$  & 10  & $ 3.49        $  &  1 \\
 HD\,69056   & $5.67\pm 0.13$  &  9  & $ 6.42\pm 0.20$  &  9  & $ 3.36\pm 0.19$  &  3 \\
 HD\,69809   & $5.79\pm 0.14$  &  8  & $ 6.65\pm 0.11$  &  7  & $ 3.27        $  &  1 \\
 HD\,69960   & $5.80\pm 0.09$  &  7  & $ 6.58\pm 0.19$  &  9  & $ 3.29        $  &  1 \\
 HD\,73226   & $5.65\pm 0.15$  &  9  & $ 6.37\pm 0.21$  &  8  & $ 3.14        $  &  1 \\
 HD\,73933   & $5.71\pm 0.14$  &  4  & $ 6.31\pm 0.22$  &  8  & $ 3.34        $  &  1 \\
 HD\,76909   & $5.81\pm 0.15$  &  7  & $ 6.64\pm 0.22$  &  9  & $ 3.26        $  &  1 \\
 HD\,77519   & $5.70\pm 0.13$  &  6  & $ 6.49\pm 0.19$  &  8  & $ 3.54        $  &  1 \\
 HD\,82943   & $5.74\pm 0.13$  &  8  & $ 6.49\pm 0.16$  &  9  & $ 3.42        $  &  1 \\
 HD\,85301   & $5.67\pm 0.14$  &  9  & $ 6.32\pm 0.25$  &  8  & $             $  &  0 \\
 HD\,87359   & $5.57\pm 0.13$  &  9  & $ 6.23\pm 0.21$  &  9  & $ 3.35\pm 0.27$  &  2 \\
 HD\,87836   & $5.75\pm 0.11$  &  7  & $ 6.63\pm 0.27$  &  8  & $ 3.23        $  &  1 \\
 HD\,90681   & $5.75\pm 0.13$  &  8  & $ 6.52\pm 0.21$  &  8  & $ 3.33        $  &  1 \\
 HD\,90722   & $5.80\pm 0.13$  &  9  & $ 6.65\pm 0.22$  &  6  & $             $  &  0 \\
 HD\,92788   & $5.77\pm 0.14$  &  9  & $ 6.60\pm 0.15$  &  8  & $ 3.17        $  &  1 \\
 HD\,97645   & $5.72\pm 0.19$  &  3  & $ 6.53\pm 0.22$  &  9  & $ 3.52        $  &  1 \\
 HD\,98618   & $5.50\pm 0.13$  &  7  & $ 6.20\pm 0.18$  &  8  & $ 3.29\pm 0.18$  &  2 \\
 HD\,98736   & $5.86\pm 0.08$  &  6  & $ 6.79\pm 0.26$  &  8  & $ 3.45\pm 0.03$  &  2 \\
 HD\,99491   & $5.91\pm 0.14$  &  9  & $ 6.63\pm 0.20$  &  8  & $ 3.44\pm 0.18$  &  2 \\
 HD\,99492   & $5.93\pm 0.07$  &  7  & $ 6.70\pm 0.18$  &  7  & $ 3.53\pm 0.04$  &  2 \\
 HD\,100069  & $5.52\pm 0.09$  &  8  & $ 6.35\pm 0.19$  &  8  & $ 3.22        $  &  1 \\
 HD\,105631  & $5.69\pm 0.11$  &  9  & $ 6.39\pm 0.27$  & 10  & $ 3.17        $  &  1 \\
 HD\,106116  & $5.63\pm 0.13$  &  8  & $ 6.39\pm 0.26$  &  9  & $ 3.22        $  &  1 \\
 HD\,106156  & $5.74\pm 0.08$  &  8  & $ 6.50\pm 0.25$  &  9  & $ 3.18        $  &  1 \\
 HD\,108942  & $5.79\pm 0.13$  &  7  & $ 6.62\pm 0.21$  &  9  & $ 3.42        $  &  1 \\
 HD\,114174  & $5.55\pm 0.09$  &  7  & $ 6.19\pm 0.20$  &  8  & $             $  &  0 \\
 HD\,116321  & $5.73\pm 0.19$  &  3  & $ 6.49\pm 0.18$  &  9  & $ 3.51        $  &  1 \\
 HD\,136618  & $5.71\pm 0.13$  &  7  & $ 6.56\pm 0.20$  &  9  & $ 3.47        $  &  1 \\
 HD\,145675  & $5.95\pm 0.13$  &  8  & $ 6.75\pm 0.28$  &  8  & $             $  &  0 \\
 HD\,147231  & $5.50\pm 0.10$  &  8  & $ 6.29\pm 0.17$  &  8  & $ 3.23\pm 0.13$  &  3 \\
 HD\,159222  & $5.58\pm 0.13$  &  8  & $ 6.41\pm 0.22$  &  8  & $ 3.23        $  &  1 \\
 HD\,190360  & $5.66\pm 0.15$  &  9  & $ 6.39\pm 0.24$  &  8  & $ 3.47        $  &  1 \\
\noalign{\smallskip}
\hline
\end{tabular}
\end{center}
\end{table*}
%%%%%%%%%%%%%%%%%%%%%%%%%%%%%%%%%%%%%%%%%%%%%%%%%%%%%%%%%%%%%%%%%%%%%%%%

\end{appendix}

\end{document}